\documentclass[10pt,twocolumn,letterpaper]{article}

\usepackage{cvpr}      
\usepackage{algorithmic}
\usepackage[ruled]{algorithm2e}
\usepackage{multirow}
\usepackage{comment}
\usepackage[dvipsnames]{xcolor}

\definecolor{cvprblue}{rgb}{0.21,0.49,0.74}
\usepackage[pagebackref,breaklinks,colorlinks,citecolor=cvprblue]{hyperref}
\usepackage[none]{hyphenat}
\def\paperID{16191} 
\def\confName{CVPR}
\def\confYear{2024}

\title{ODDR: Outlier Detection \& Dimension Reduction Based Defense Against Adversarial Patches}

\author{%
Nandish Chattopadhyay$^*$$^{1}$, Amira Guesmi$^*$$^{1}$, Muhammad Abdullah Hanif$^1$, \\ 
Bassem Ouni$^2$, Muhammad Shafique$^1$ \\
$^1$ eBrain Lab, Division of Engineering, New York University (NYU) Abu Dhabi, UAE \\ $^2$ AI and Digital Science Research Center, Technology Innovation Institute (TII), Abu Dhabi, UAE\\}

\begin{document}
\maketitle
\def\thefootnote{*}\footnotetext{These authors contributed equally to this work}
\begin{abstract}
Adversarial attacks present a significant challenge to the dependable deployment of machine learning models, with patch-based attacks being particularly potent. These attacks introduce adversarial perturbations in localized regions of an image, deceiving even well-trained models. In this paper, we propose Outlier Detection and Dimension Reduction (ODDR), a comprehensive defense strategy engineered to counteract patch-based adversarial attacks through advanced statistical methodologies.
Our approach is based on the observation that input features corresponding to adversarial patches—whether naturalistic or synthetic—deviate from the intrinsic distribution of the remaining image data and can thus be identified as outliers. ODDR operates through a robust three-stage pipeline: Fragmentation, Segregation, and Neutralization. This model-agnostic framework is versatile, offering protection across various tasks, including image classification, object detection, and depth estimation, and is proved effective in both CNN-based and Transformer-based architectures.
In the Fragmentation stage, image samples are divided into smaller segments, preparing them for the Segregation stage, where advanced outlier detection techniques isolate anomalous features linked to adversarial perturbations. The Neutralization stage then applies dimension reduction techniques to these outliers, effectively neutralizing the adversarial impact while preserving critical information for the machine learning task.
Extensive evaluation on benchmark datasets against state-of-the-art adversarial patches underscores the efficacy of ODDR. Our method enhances model accuracy from 39.26\% to 79.1\% under the GoogleAp attack, outperforming leading defenses such as LGS (53.86\%), Jujutsu (60\%), and Jedi (64.34\%).

\end{abstract}    
\section{Introduction \& Related Work}
\label{sec:intro}
Adversarial attacks represent a serious threat to the robustness and performance of deep neural network (DNN) models \cite{Goodfellow2015ExplainingAH}. These attacks involve the strategic introduction of adversarial perturbations into test samples, leading to significant disruptions in the model's predictions. A particularly effective form of these attacks is the insertion of localized patches into test images, which exploits a critical vulnerability. This allows attackers to force DNN models into making errors in essential tasks such as image classification, object detection, and depth estimation \cite{Hu21, guesmi2024ssap}.

\begin{figure}[!htbp] 
\centerline{\includegraphics[width=\columnwidth]{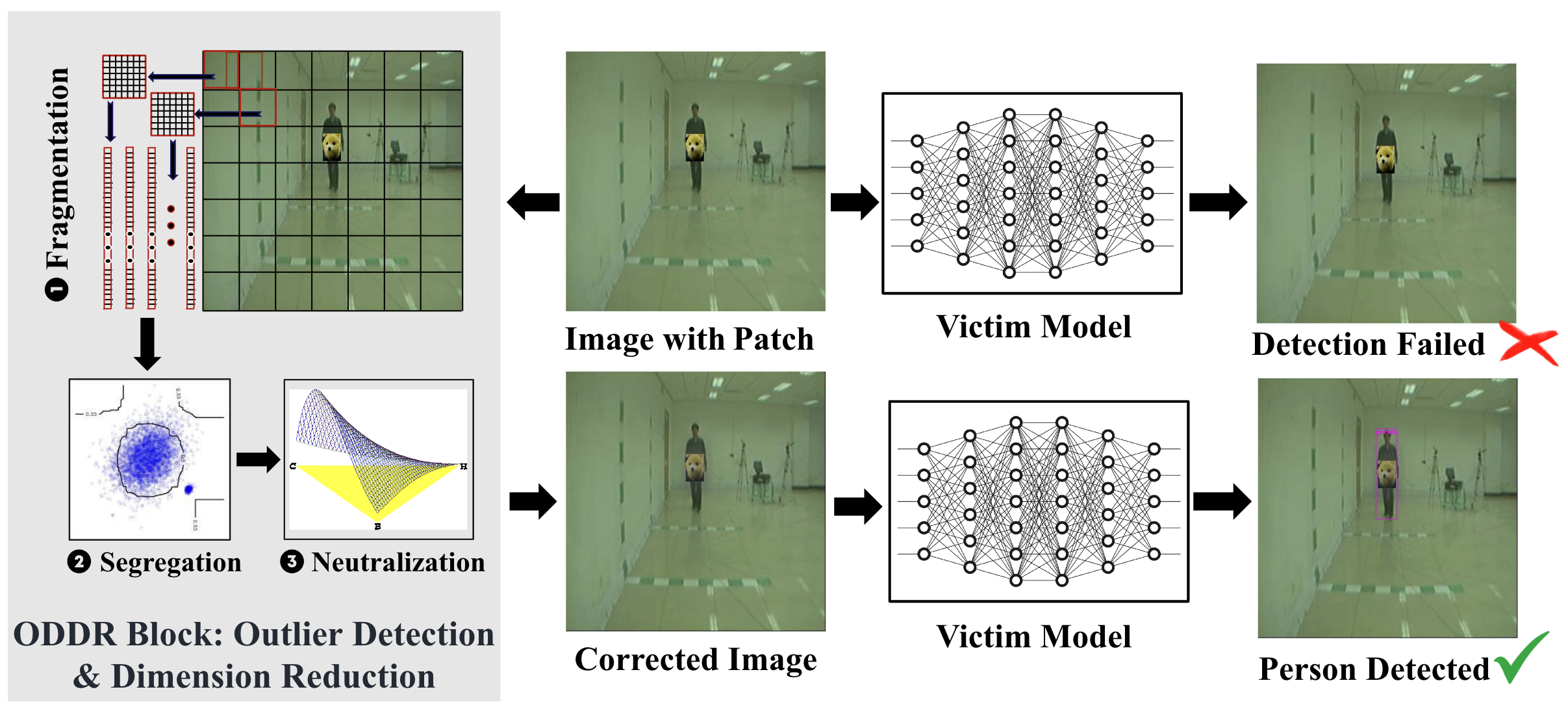}} 
\caption{Overview of the Proposed ODDR Defense Methodology: The three-stage pipeline—Fragmentation, Segregation, and Neutralization—demonstrating the process of identifying and mitigating adversarial patches in the input image features.}
\label{fig:concept}
\end{figure}
Patch-based attacks are a widely recognized and practical form of adversarial attack, particularly valued for their adaptability in scenarios with limited access \cite{guesmi2023physical}. These attacks function much like discreet stickers, easily applied in real-world situations where adversaries may face constraints in resources or access. The subtlety of patch-based attacks makes them especially elusive, underscoring the critical need for robust defense mechanisms.

There are two primary approaches to defend against patch-based attacks: \textit{Empirical defenses} and \textit{Certified defenses}.

\noindent\textbf{Empirical defenses} include methods like Localized Gradient Smoothing (LGS) \cite{naseer2019local}, which normalizes gradient values and uses a moving window to identify high-density regions based on specific thresholds. Jujutsu \cite{Jujutsu} focuses on localizing adversarial patches and differentiating them from benign samples by leveraging generative adversarial networks to reconstruct ‘clean’ images from adversarial examples. Jedi \cite{jedi} utilizes input entropy analysis and image inpainting to recover the original image.

\noindent\textbf{Certified defenses} aim to provide provable robustness against attacks. For instance, De-randomized Smoothing (DS) \cite{levine2020randomized} builds a certified defense by creating a smoothed classifier through ensembling local predictions on pixel patches. PatchGuard (PG) \cite{xiang2021patchguard} enhances robustness by enforcing a small receptive field within deep neural networks and employing secure feature aggregation.

Despite their effectiveness, these defenses are prone to generating false positives \cite{naseer2019local} and often struggle to accurately distinguish between adversarial and clean samples. Additionally, some defenses may inadvertently remove or alter important features \cite{xiang2021patchguard, levine2020randomized}, leading to degraded model performance even on benign samples.

For the second aspect of the problem, having identified the region that contains the adversarial patch, rather than relying on intricate and computationally intensive neural network-based patch neutralization techniques, we advocate for the use of simple Dimension Reduction \cite{freedman}. Through matrix-based transformations, this approach proves adept at neutralizing the impact of the patch without loss of surrounding pertinent information. This information is often critical for the underlying machine learning task. 

\noindent\textbf{Novel Contributions -- } The main contributions of this paper are: 
\begin{itemize}
    \item We introduce a novel, model-agnostic defense method, Outlier Detection and Dimension Reduction (ODDR), which accurately identifies and localizes adversarial patches by detecting clusters of outliers within input features. By applying targeted dimension reduction over a window surrounding the detected outlier cluster, ODDR effectively neutralizes the adversarial patch while preserving crucial information. 
    \item We demonstrate the efficacy and adaptability of ODDR across various tasks, including image classification, object detection, and monocular depth estimation. Our defense achieves a robust accuracy of 79.1\% against the GoogleAP attack on the ImageNet dataset using CNN-based architectures and 83.1\% using Transformer-based architectures.
    \item For object detection tasks, our defense mechanism achieves robust average precision scores of 93.54\% and 83.21\% when countering the AdvYOLO attack on the CASIA and INRIA datasets, respectively.
    \item In depth estimation, our defense significantly reduces the mean depth estimation error from 0.59 to 0.1 when mitigating the impact of SSAP-based attacks, demonstrating its effectiveness in preserving model accuracy.
    \item Our proposed approach improves the accuracy from 39.26\% to 79.1\% when performing GoogleAp attack, which surpasses that of the prominent state-of-the-art like LGS (53.86\%), Jujutsu (60\%), and Jedi (64.34\%).
\end{itemize}
\section{Theoretical Underpinning}
\label{theory}
In this section, we delve into the foundational theoretical principles that constitute the building blocks of the proposed ODDR scheme. We outline the two components of the defense mechanism.

\subsection{Segregating Input Features}
We use the Isolation Forest algorithm as a robust tool for detecting outliers and anomalies within datasets. Operating under the paradigm of Unsupervised Learning, this technique excels in scenarios where labeled data is unavailable, making it particularly versatile in diverse applications. It uses Binary Decision Trees through a bagging approach, drawing parallels to the well-known Random Forest technique employed in supervised learning. The core hypothesis guiding Isolation Forest is predicated on the assumption that anomalies represent a minority within the dataset, comprising fewer instances than their normal counterparts. Additionally, these anomalies are posited to possess attribute-values that diverge from those associated with normal instances. In simpler terms, anomalies are characterized as being both fewer in number and different from the rest of the samples \cite{IF}. This particularly makes it a good fit for detecting adversarial patches from input features of the test samples.
Leveraging these assumptions, Isolation Forest excels in isolating anomalies by constructing an ensemble of Binary Trees for a given dataset. Anomalies, given their distinctive nature, tend to exhibit shorter paths within these trees compared to normal instances \cite{IFextended}.

\subsection{Outlier Detection}
Let's formally define the mechanism of constructing an isolation tree for the purpose of detecting outliers. Consider a dataset $X = \{ x_1, \ldots, x_n \}$, comprising $n$ vectors representing instances from a multivariate distribution of dimension $d$.  
To build the isolation tree, the dataset $X$ undergoes a recursive division process based on the random selection of an attribute $q$ and a corresponding threshold value $p$. This process continues until one of the following three conditions is met:

\begin{itemize}
    \item the isolation tree attains maximum permissible height
    \item $|X| = 1$
    \item the data in the dataset $X$ has equal values
\end{itemize}

The objective of anomaly or outlier detection can be achieved if one can deduce a way to rank the samples reflecting degree of anomaly. This can be done by running a sorting algorithm on the path lengths or anomaly scores. This naturally necessitates the definition of the anomaly scores. The path length $h(x)$ for a sample point $x$ is defined as the count of nodes that $x$ traverses in the isolation tree from the root node to the conclusion of the traversal at an external node. Drawing on the structural similarities between isolation trees and Binary Search Trees, it can be inferred that the estimated average path length aligns with that of an unsuccessful search in a Binary Search Tree. Well-established in the literature, it is known that for a dataset comprising $n$ samples, the average path length of an unsuccessful search is:
\begin{equation}
\label{path}
    c(n) = 2H(n-1)-(2(n-1)/n), 
\end{equation}
with $H(i)$ denoting the Harmonic number, estimated as $ln(i) + 0.57721567$ [Euler's constant]. Additionally, $c(n)$ represents the mean of $h(x)$ over $n$. Consequently, the anomaly score $s$ can be defined as:

\begin{equation}
\label{score}
    Score(x) = s(x,n) = 2^{-\frac{E(h(x))}{c(n)}}
\end{equation}

where $E(h(x))$ represents the mean of $h(x)$ derived from the isolation forest, which is constructed from multiple isolation trees. It follows that when $E(h(x)) \to c(n)$, $s \to 0.5$; when $E(h(x)) \to 0$, $s \to 1$; and when $E(h(x)) \to n-1$, $s \to 0$. 
The interpretation of the anomaly score $s$ is that if the samples possess an anomaly score close to unity, they are highly likely to be outliers and those with a score much smaller than $0.5$ are likely to belong to the distribution. If a significant portion of samples returns a score around the $0.5$ mark, it is highly probable that there are not many outliers in the dataset. Anomalies are identified by their shorter path lengths within the tree structure, with path length denoting the number of edges traversed from the root node.

\subsection{Dimension Reduction}
Dimension Reduction has been successfully used as a method of mitigating adversarial attacks \cite{cod_1}. This technique maps the input into a lower dimensional space, preserving a chosen amount of variability, thereby getting rid of significant proportion of the adversarial noise \cite{cod_2}.  Intuitively, this projects the matrix on to an orthonormal basis and the variability within it is expressed as a linear combination of the components \cite{freedman}. 
At first, if we consider that an image (a coloured image has three channels and therefore three such matrices) is a $m$x$n$-matrix $M$, then one can use Singular Value Decomposition to factorize the same into $M = U \Sigma V^{T}$, where $U$ and $V$ are orthogonal matrices and $\Sigma = diag(\sigma_{1}, \ldots, \sigma_{r})$, where $r=min(m,n)$ such that $\sigma_{1} \geq \ldots \geq \sigma_{r} \geq 0$. In this case, the singular values are the $\sigma_{i}$s, and thereby we have the top $r$ columns of $V$ and $U$ being the right and left singular vectors respectively. This formulation is necessary for dimensionality reduction using Truncated Singular Value Decomposition as explained hereafter. 
For the matrix $M$ with a rank $r$, one can represent $M = U_{r} \Sigma_{r} V_{r}^{T}$ where $U_{r}$ and $V_{r}$ comprises of the left and right singular vectors. For any $k$, such that $(k <r)$, one can therefore obtain $M_{k}$, where $M_{k} = U_{k} \Sigma_{k} V^{T}_{k}$, which is lower rank approximation of the matrix. This is used in dimensionality reduction, facilitated by the fact that the singular values are so arranged such that they provide the ordering of the information contained in such truncated approximations.  
Specifically, in our implementation, we have used the proportion of the singular values as a surrogate to estimate the amount of information preserved upon the image mask $M$ being projected to a lower dimension. We have denoted this information preservation parameter as $inf$ and used it as a tuning parameter for our defense mechanism.
\section{ODDR Defense Mechanism}
\label{method}
As illustrated in Figure \ref{fig:concept}, the proposed ODDR defense technique encompasses three stages. The initial stage involves splitting the image into kernels using a moving window. This fragmentation process deconstructs the image into equally sized chunks, the dimensions of which are fine-tuned as hyper-parameters for specific applications. This stage is referred to as Fragmentation. Subsequently, the obtained fragments are flattened into vectors, and the outlier detection task is executed on these vectors, denoted as Feature Segregation. Finally, Dimension Reduction is applied to a window positioned on the cluster of fragments identified as outliers. This step aims to neutralize the adversarial patch without compromising information in the surrounding vicinity. The entire algorithm is detailed in Algorithm \ref{alg1}, with each component and function elucidated in the subsequent subsections.
\begin{algorithm}
\caption{ODDR: Outlier Detection \& Dimension Reduction}
\label{alg1}
$\boldsymbol{IN}:$ $I$: original image, $k$: kernel size, $str$: stride length, $c$: confidence of anomalies, $T$: number of trees in isolation forest, $d$: size of square mask, $inf$: information preserved after SVD, $minPts$: minimum fragments in an anomaly cluster\\
$\boldsymbol{OUT}:$ $I'$: Image with neutralised patch \\
/*$\boldsymbol{Fragmentation}$*/\\
Generate $n$ fragments $ X \gets (x_1, \ldots, x_n)$ from image $I$, using kernel size = $k$ and stride = $str$ \\
/*$\boldsymbol{Segreagation}$*/\\
$\textbf{Set}$ $s=0.3$ $\times$ $size(X)$ \\
$iForest = IsolationForest(X,T,s)$ (From Algo 2 in the Appendix) 
\\
$\boldsymbol{initialize}$ $ Y = (y_1, \ldots, y_n)$ \\
$\textbf{for}$ $i=1$ to $n$ do: \\
$y_{i} = Score(x_{i})$, where $x_{i} \in X$  (From Eq. \ref{score})\\
$\textbf{end}$ $\textbf{for}$ \\
Sort $Y$ in descending order \\
Select $(1-c)*n$ top $y_{i}$s as anomalies $(z_1, \ldots, z_{r}) \to Z$ \\
Select clusters $(C_1, \ldots, C_p)$ in $Z$ such that $size(C_i) \geq minPts$\\
/*$\boldsymbol{Neutralization}$*/\\
Calculate cluster centers $z\_cen_{C_i}$ and masks $M_{C_i}$ of size $d$ as: \\
$\textbf{for}$ $i=1$ to $p$ do: \\
Cluster center $z\_cen_{C_i}$ = $max(z_i)$ $\forall$  $z_i \in C_i$\\
Create masks $M_{C_i}$ from loc($z\_cen_{C_i}-(d/2))$ to loc($z\_cen_{C_i}+(d/2))$ \\
Perform Dimension Reduction using Singular Value Decomposition as: \\
$\textbf{for}$ $i=1$ to $p$ do: \\
$M'_{C_i}$ $\gets$ $SVD(M_{C_i}, inf)$ \\
Superimpose $M'_{C_i}$s in place of $M_{C_i}$s in image $I$ to get $I'$ \\
$\boldsymbol{return}$ $I'$
\end{algorithm}

\subsection{Fragmentation}
The initial step in processing the input image is Fragmentation. This involves segmenting the image into partially overlapping fragments using a moving square kernel. The size of this kernel, along with the stride length determining the movement for each new fragment, is a system hyper-parameter fine-tuned for optimal performance. Upon generating the fragments, each with three channels and a square shape, the data is flattened into long vectors. This flattening is a crucial preparatory step for the subsequent stage in the process, Feature Segregation.

\subsection{Segregation}
The notion of segregation of anomalous input features within the feature space is carried out as an outlier detection exercise. We propose that an ensemble of isolation trees, which is an isolation forest, is best suited for this purpose. The fragments generated using the aforementioned Fragmentation process serves as the dataset for the outlier detection algorithm. The anomaly detection using Isolation Forests is outlined in Algorithm 2 in the Appendix, which is essentially an ensemble of Isolation Trees which are generated using Algorithm 3.  
Once the trees are generated, the path lengths for the leaf nodes are calculated using the Equation \ref{path}, which is thereafter used to estimate the anomaly score using the Equation \ref{score}, as described earlier.  
Once the anomaly scores ($y_i$) are ascertained to each fragment ($x_i$), the top outliers are selected ($Z$) based on a pre-defined threshold ($c$), which is set as a hyper-parameter. Therefore, $Z$ contains fragments $z_{i}$s, detected as outliers. Since Fragmentation is a deterministic process, using $k$ and $str$, we know the placement of every fragment $x_{i}$ , and so for every $z_{i} \in Z$, we check the overlapping fragments to look for the clusters ($C_i$s), such that the size of each cluster is greater than a pre-defined hyper-parameter $minPts$.

\subsection{Neutralization}
For every detected cluster \mbox{$(C_1, \ldots, C_p)$}, the cluster center of each is ascertained to be the fragment with the highest anomaly score. Considering the central pixel of that particular fragment as the center, the square masks are generated of size $d$, extending from the location \mbox{[$z\_cen_{C_i}-(d/2)$]} to \mbox{[$z\_cen_{C_i}+(d/2)$]} on either side of the central point. 
The mask derived from the Segregation process encompasses the entirety or, at the very least, the majority of the adversarial patch, whether naturalistic or otherwise. This assertion is grounded in the fact that the mask is chosen from the region where the cluster of anomalies or outlier fragments is situated. \textit{Comprehensive testing and validation of this approach are presented in the Experiments Section later in the paper.}
 
To render the detected adversarial patch ineffective, two techniques have been explored. The first approach, though sub-optimal, involves replacing each pixel in the mask with the mean value of all the pixels within the region covered by the mask. This method performs effectively when the mask aligns closely with the adversarial patch. However, for a broader region, it may result in a potential loss of valuable information in areas where the mask does not overlap with the adversarial patch. To solve this problem, we propose Dimension Reduction. We apply a singular value decomposition on the mask, with a specific hyper-parameter that selects the proportion of information to preserve (corresponding to the number of singular values). The details of this are explained in the previous section. Either way, both of these techniques are computationally much simpler than the use of neural network based in-painting that is used in many defense techniques available in the literature \cite{jedi}. 
\section{Experimental Results}
\label{results}
In this section, we conduct a thorough evaluation of the effectiveness of our proposed defense mechanism against various patch-based attacks across different classification, object detection, and depth estimation tasks. 
\subsection{Experimental Setup}
We assess ODDR performance across diverse settings commonly used in computer vision applications:\\
\textbf{Task: Classification - }\\
\textbf{Datasets:} We use the ImageNet dataset \cite{imagenet} for its extensive class variety and the Caltech-101 dataset \cite{caltech} for specific object recognition tasks.\\
\textbf{Models:} Our assessment uses ResNet-152 \cite{he2016deep}, ResNet-50 \cite{he2016deep}, VGG-19 \cite{simonyan2014very}, Inception V3 \cite{inception}, Vision Transformers \cite{VIT} and Swin Transformers \cite{swin} for ImageNet and Caltech-101 classification tasks.\\
\textbf{Task: Detection - } \\
\textbf{Datasets:} For outdoor detection challenges, we use the INRIA dataset \cite{inria}, capturing uncontrolled environmental variations and includes instances of multiple patches. For indoor scenarios, we use the CASIA dataset \cite{casia}.\\
\textbf{Models:} The YOLO model \cite{yolov4} is employed for detection tasks. We test both outdoor and indoor scenarios.\\
\textbf{Task: Monocular Depth Estimation (MDE) - }\\
\textbf{Datasets:} For monocular depth estimation challenges, we use the KITTI dataset \cite{kitti}. The dataset encompasses a wide variety of road types, including local streets, rural roads, and highways, providing a robust and diverse set of scenarios for depth estimation tasks.\\
\textbf{Models:} We employ the MIMdepth model \cite{mimdepth} for monocular depth estimation tasks. MIMdepth utilizes a self-supervised learning approach, combining masked image modeling (MIM) with a transformer architecture to enhance the accuracy and robustness of monocular depth estimation.

\noindent \textbf{Threat Model:}
We operate under the strongest assumption, a white-box scenario, where the attacker possesses complete knowledge of the victim DNN, including its architecture and parameters. As in other proposed defenses \cite{levine2020randomized, xiang2021patchguard, naseer2019local}, the adversary is capable of substituting a specific region of an image with an adversarial patch, with the goal of consistently inducing misclassification/misdetection/misestimaion across all input instances.

\noindent \textbf{Patch-based Attacks:}
In this evaluation, we employ seven state-of-the-art adversarial patches to rigorously assess model performance. For classification tasks, we utilize the Adversarial Patch (GAP) \cite{googleap}, LAVAN \cite{lavan}, Generative Dynamic Patch Attack (GDPA) \cite{li2021generative}, and Shape Matters (SM) \cite{chen2022shape}. For detection tasks, we apply the AdvYOLO adversarial patch \cite{thys2019} and the Naturalistic Patch \cite{Hu21}. Additionally, to evaluate Transformer-based monocular depth estimation (MDE) models, we use the shape-sensitive adversarial patch (SSAP) \cite{guesmi2024ssap}. Further details in the Appendix.
These patches represent cutting-edge techniques in the domain, ensuring a robust and comprehensive assessment of the ODDR defense mechanism across different challenges.

\subsection{Results}
Throughout the experiments, for our defense, we used a kernel size of $20$x$20$ pixels, with a stride length of $10$ for Fragmentation. For Segregation we varied the confidence of anomalies $c$ between $0.8-0.85$ and used $500$ trees in the isolation forest. For the Dimension Reduction using SVD, the information preservation parameter is varied between $70-85\%$. 

\subsubsection{ODDR in Classification Benchmarks:}
In our evaluation, the primary focus was on assessing the model's robust accuracy under adversarial conditions. To demonstrate the effectiveness of our defense strategy, we first generated adversarial patches using four distinct attack methods: LAVAN, GoogleAP, GDPA, and Shape Matters (SM). We then measured the model's robust accuracy across varying patch sizes and architectures, including both CNN and transformer-based models. Table \ref{tab1} showcases the outstanding performance of our defense technique, which significantly enhances robust accuracy against the aforementioned attacks on the ImageNet and CalTech datasets, with minimal impact on baseline accuracy. Notably, our defense achieves robust accuracy rates of 79.1\% against GoogleAP and 74.1\% against LAVAN, underscoring its effectiveness in mitigating these adversarial threats.
\begin{table}[h]
\resizebox{\columnwidth}{!}{%
\begin{tabular}{lccccccccc}
\hline
\multicolumn{1}{c|}{IMAGENET} & \multicolumn{1}{c|}{\multirow{2}{*}{Clean}} & \multicolumn{2}{c|}{GAP} & \multicolumn{2}{c|}{LAVAN} & \multicolumn{2}{c|}{GDPA} & \multicolumn{2}{c}{SM} \\ \cline{1-1} \cline{3-10} 
\multicolumn{1}{c|}{Model} & \multicolumn{1}{c|}{} & Adv & \multicolumn{1}{c|}{ODDR} & Adv & \multicolumn{1}{c|}{ODDR} & Adv & \multicolumn{1}{c|}{ODDR} & Adv & ODDR \\ \hline \hline 
\multicolumn{1}{l|}{ResNet 152} & \multicolumn{1}{c|}{81.2} & 39.3 & \multicolumn{1}{c|}{79.1} & 10.1 & \multicolumn{1}{c|}{74.1} & 56.4 & \multicolumn{1}{c|}{73.1} & 15.1 & 78.3 \\
\multicolumn{1}{l|}{ResNet 50} & \multicolumn{1}{c|}{78.4} & 38.8 & \multicolumn{1}{c|}{75.6} & 10.2 & \multicolumn{1}{c|}{70.2} & 55.8 & \multicolumn{1}{c|}{72.6} & 16.4 & 72.8 \\
\multicolumn{1}{l|}{VGG 19} & \multicolumn{1}{c|}{74.2} & 39.1 & \multicolumn{1}{c|}{72.8} & 11.1 & \multicolumn{1}{c|}{71.1} & 60.1 & \multicolumn{1}{c|}{71.8} & 17.1 & 71.9 \\
\multicolumn{1}{l|}{Inception v3} & \multicolumn{1}{c|}{79.8} & 38.2 & \multicolumn{1}{c|}{77.6} & 10.9 & \multicolumn{1}{c|}{77.1} & 62.1 & \multicolumn{1}{c|}{72.9} & 12.8 & 74.1 \\
\multicolumn{1}{l|}{ViT\_b\_16} & \multicolumn{1}{c|}{85.8} & 36.7 & \multicolumn{1}{c|}{83.1} & 9.2 & \multicolumn{1}{c|}{82.9} & 68.4 & \multicolumn{1}{c|}{82.6} & 11.9 & 82.3 \\
\multicolumn{1}{l|}{swin\_v2\_b} & \multicolumn{1}{c|}{84.5} & 35.9 & \multicolumn{1}{c|}{81.9} & 6.9 & \multicolumn{1}{c|}{81.2} & 69.8 & \multicolumn{1}{c|}{81.8} & 13.4 & 80.8 \\ \hline
 &  &  &  &  &  &  &  &  &  \\ \hline
\multicolumn{1}{c|}{CALTECH-101} & \multicolumn{1}{c|}{\multirow{2}{*}{Clean}} & \multicolumn{2}{c|}{GAP} & \multicolumn{2}{c|}{LAVAN} & \multicolumn{2}{c|}{GDPA} & \multicolumn{2}{c}{SM} \\ \cline{1-1} \cline{3-10} 
\multicolumn{1}{c|}{Model} & \multicolumn{1}{c|}{} & Adv & \multicolumn{1}{c|}{ODDR} & Adv & \multicolumn{1}{c|}{ODDR} & Adv & \multicolumn{1}{c|}{ODDR} & Adv & ODDR \\ \hline \hline 
\multicolumn{1}{l|}{ResNet 152} & \multicolumn{1}{c|}{94.1} & 48.6 & \multicolumn{1}{c|}{90.8} & 15.6 & \multicolumn{1}{c|}{91.1} & 45.7 & \multicolumn{1}{c|}{88.4} & 23.6 & 91.8 \\
\multicolumn{1}{l|}{ResNet 50} & \multicolumn{1}{c|}{90.9} & 49.2 & \multicolumn{1}{c|}{86.4} & 17.1 & \multicolumn{1}{c|}{87.3} & 47.9 & \multicolumn{1}{c|}{87.1} & 25.9 & 86.3 \\
\multicolumn{1}{l|}{VGG 19} & \multicolumn{1}{c|}{88.6} & 47.1 & \multicolumn{1}{c|}{85.6} & 15.3 & \multicolumn{1}{c|}{84.9} & 46.1 & \multicolumn{1}{c|}{85.2} & 21.4 & 86.6 \\
\multicolumn{1}{l|}{Inception v3} & \multicolumn{1}{c|}{89.4} & 48.1 & \multicolumn{1}{c|}{86.5} & 18.2 & \multicolumn{1}{c|}{85.4} & 49.2 & \multicolumn{1}{c|}{86.8} & 19.4 & 84.1 \\
\multicolumn{1}{l|}{ViT\_b\_16} & \multicolumn{1}{c|}{95.3} & 45.2 & \multicolumn{1}{c|}{91.8} & 13.8 & \multicolumn{1}{c|}{92.1} & 51.2 & \multicolumn{1}{c|}{91.9} & 25.1 & 92.3 \\
\multicolumn{1}{l|}{swin\_v2\_b} & \multicolumn{1}{c|}{94.9} & 46.3 & \multicolumn{1}{c|}{90.7} & 17.9 & \multicolumn{1}{c|}{90.3} & 48.3 & \multicolumn{1}{c|}{92.1} & 20.8 & 92.7 \\ \hline
\end{tabular}
}
\caption{ODDR robustness for Image Classification task.}
\label{tab1}
\end{table}

\subsubsection{ODDR in Detection Benchmarks:}
In the context of detection tasks, we evaluated the Robust Average Precision and Recovery Rate against the AdvYOLO attack on the INRIA and CASIA datasets, as presented in Table \ref{tab:detection}. The Recovery Rate measures the proportion of successfully recovered outputs by the defense relative to the total number of successful attacks, providing a clear indication of the defense mechanism's effectiveness. Our defense technique demonstrates exceptional performance, achieving a robust average precision of 93.54\% on the INRIA dataset and 83.21\% on the CASIA dataset.

\begin{table}[!htbp]
\resizebox{\columnwidth}{!}{%
    \centering
    \begin{tabular}{c|c|c|c}
     \hline
      Dataset  & Clean Accuracy & Adversarial Accuracy & Robust Average Precision \\
     \hline \hline 
     INRIA     &  96.42\%   &  39.28\%   &   83.21\%  \\ 
     CASIA     &  96.77\%   &   32.25\%  &  93.54\%   \\
      \hline
    \end{tabular}
    }
    \caption{Performance of ODDR against AdvYOLO-based attack.}
    \label{tab:detection}
\end{table}

\subsubsection{ODDR in Depth Estimation Benchmarks:}

For depth estimation tasks, we assessed the effectiveness of our defense using three key metrics, as defined in \cite{guesmi2024ssap}: mean depth estimation error ($E_{d}$), the ratio of the affected region ($R_{a}$), and mean square error (MSE). Further details on these metrics can be found in the Appendix. As demonstrated in Table \ref{tab:depth}, our ODDR technique significantly mitigates the impact of the SSAP patches, reducing the mean depth estimation error from 0.59 to 0.1 and shrinking the affected region from 99\% to 21\%, thereby effectively preserving the accuracy of depth estimation.
\begin{table}[!htbp]
\centering
\resizebox{0.4\columnwidth}{!}{%
    \begin{tabular}{l|c|c|c}
    \hline
    KITTI & \textbf{$E_{d}$}  & \textbf{$R_{a}$}  & \textbf{$MSE$}\\
    \hline
      SSAP         &  0.59 & 0.99  & 0.53\\
      ODDR     &  0.11 & 0.21 &  0.04\\
     \hline
  \end{tabular} }
  \captionof{table}{\label{tab:depth}Performance of ODDR against SSAP-based attack.}
\end{table}

\begin{figure}[!htbp]
\centerline{\includegraphics[width=\columnwidth]{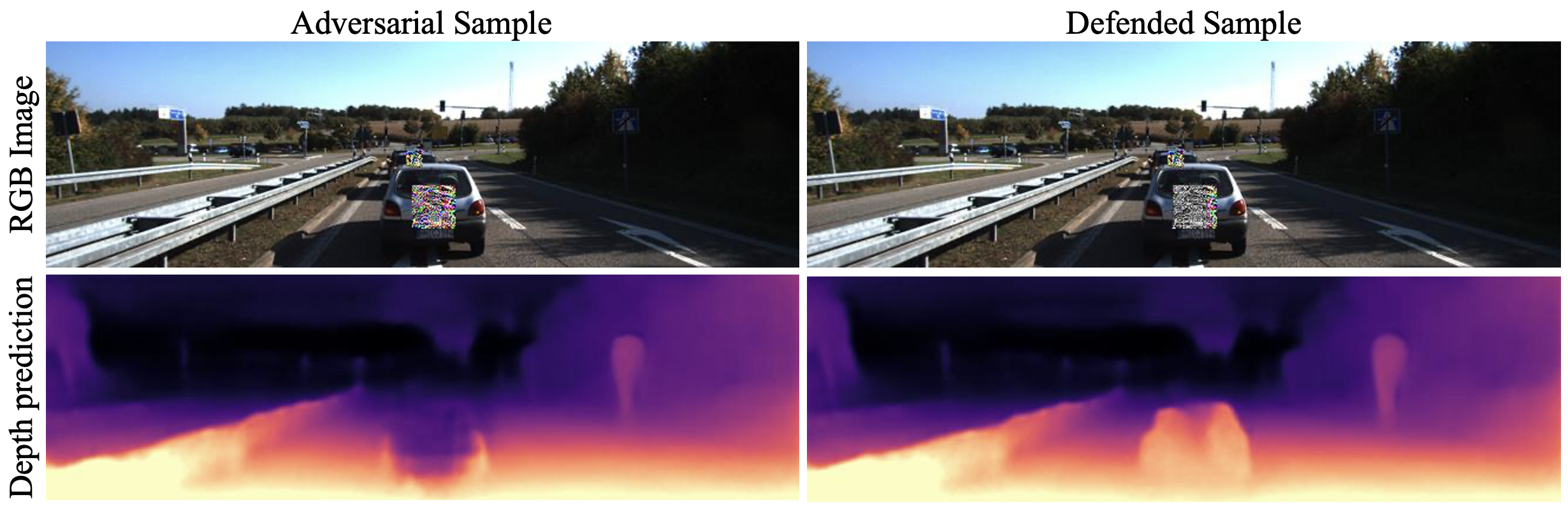}} 
\caption{Illustration on the impact of ODDR on depth SSAP-based attack \cite{guesmi2024ssap}.} 
\label{fig:depth}
\end{figure}

\subsection{ODDR vs State-of-the-art}

We conducted a comprehensive comparative analysis of ODDR against several state-of-the-art defenses, including LGS \cite{naseer2019local}, Jujutsu \cite{Jujutsu}, and Jedi \cite{jedi}, across all previously outlined experiments. For further context, we also compared ODDR with two certified defenses: De-randomized Smoothing \cite{levine2020randomized} and PatchGuard \cite{xiang2021patchguard}. As shown in Table \ref{tab2}, our defense technique outperforms existing methods, achieving a robust accuracy of 79.1\%. This surpasses the robust accuracy achieved by LGS (53.86\%), Jujutsu (60\%), and Jedi (64.34\%). Additionally, we compared ODDR with a Dimension Reduction-based defense \cite{defensivedr}, which achieved an accuracy of 66.2\%, further highlighting the superior performance of our approach.

\begin{figure}[!htbp]
\centerline{\includegraphics[width=\columnwidth]{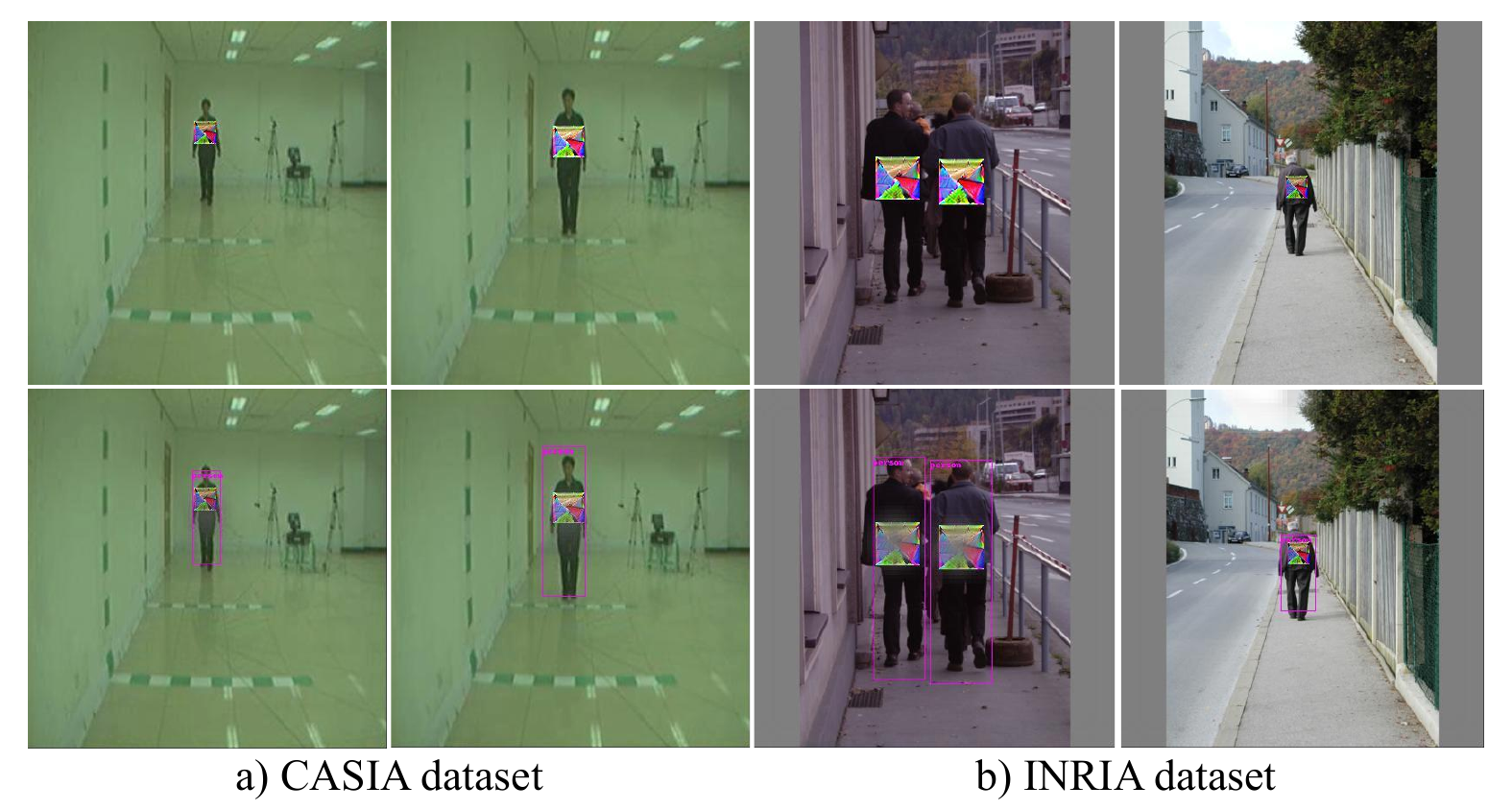}} 
\caption{Illustration of the impact of ODDR on AdvYOLO patch-based attacks \mbox{\cite{thys2019}}: \textit{Top row}: Adversarial images. \textit{Bottom row}: Defended images for two datasets: a) CASIA representing indoor settings, and b) INRIA representing outdoor settings.}
\label{fig:_res__yolo}
\end{figure}

In addition, we compared our defense with two certified defenses to quantify the trade-off between ensuring provable robustness and overall performance, despite the inherent unfairness of this comparison. As expected, certified adversarial patch defenses demonstrated lower performance compared to empirical defenses, with De-randomized Smoothing \cite{levine2020randomized} achieving 35.02\% robust accuracy and PatchGuard \cite{xiang2021patchguard} achieving 30.96\% robust accuracy. These results highlight the significant trade-off between provable robustness and empirical utility, underscoring the practical limitations of certified defenses in real-world scenarios.

\begin{table}[h]
\centering
\resizebox{0.5\textwidth}{!}{%
\begin{tabular}{l|c|cc|cccc}
\hline
\multicolumn{1}{c|}{Model} & Attack & Clean & Adv & LGS & Jedi & DR & ODDR \\ \hline \hline 
\multirow{3}{*}{ResNet 152} & GAP & 81.2 & 39.3 & 53.9 & 64.4 & 66.2 & 79.1 \\
 & LAVAN & 81.2 & 10.1 & 49.9 & 54.6 & 59.4 & 74.1 \\
 & GDPA & 81.2 & 56.4 & 64.2 & 61.5 & 60.2 & 73.1 \\ \hline
\multirow{3}{*}{Inception v3} & GAP & 79.8 & 38.2 & 55.1 & 66.8 & 61.3 & 77.6 \\
 & LAVAN & 79.8 & 10.9 & 50.7 & 59.4 & 55.4 & 77.1 \\
 & GDPA & 79.8 & 62.1 & 66.9 & 71.4 & 58.6 & 72.9 \\ \hline
\multirow{3}{*}{swin\_v2\_b} & GAP & 84.5 & 35.9 & 65.2 & 70.6 & 62.6 & 81.9 \\
 & LAVAN & 84.5 & 6.9 & 60.9 & 57.8 & 53.8 & 81.2 \\
 & GDPA & 84.5 & 69.8 & 76.3 & 68.4 & 58.9 & 81.8 \\ \hline
\end{tabular}
}
\caption{ODDR comparison across attacks on ImageNet dataset.}
\label{tab2}
\end{table}

In detection tasks, as demonstrated in Table \ref{tab:detection_}, our defense outperforms in terms of both robust average precision and recovery rate when deployed against the advYOLO patch, across both INRIA and CASIA datasets.

\begin{table}[!htbp]
    \resizebox{\columnwidth}{!}{%
    \centering
    \begin{tabular}{c|c|c|c|c}
    \hline
             &   \multicolumn{2}{|c|}{INRIA} & \multicolumn{2}{|c}{CASIA} \\
     Defense    & Robust Avg.  & Recovery   & Robust Avg.   & Recovery  \\
                &  Precision  &  Rate   &  Precision   &  Rate     \\
    \hline \hline 
    LGS \cite{naseer2019local}  & 21\%   &  27\%&  84\%   &   48\%\\
    Jedi \cite{jedi}          & 28.03\%  & 42\%   &  80\%  &  50\%  \\
     Ours      (ODDR)               & \textbf{82.14\%}  & \textbf{43.86\%}  & \textbf{93.54\%}   & \textbf{61.29\%}   \\
    \hline
    \end{tabular}
    }
    \caption{Performance of our proposed defense ODDR compared to two state-of-the-art defenses against AdvYOLO.}
    \label{tab:detection_}
\end{table}
\section{Discussion}
\label{disc}
In this section, we explore additional performance aspects of ODDR, including its resilience against adaptive attacks, results from the ablation study, impact on model interpretability, computational cost, and hyperparameter tuning.
\subsection{Ablation Study}
The proposed ODDR technique identifies the adversarial noise in the patches as an outlier to the data distribution of the rest of the input sample. For each component, we chose the best algorithm that maximises the overall inference accuracy. For \textit{Outlier Detection}, we found Isolation Forests to outperform clustering based techniques like DBSCAN \cite{dbscan}. The percentage gain in the overlap of the actual adversarial patch and the detected outliers cluster was up to $20\%$ (ImageNet). For \textit{Neutralization}, we used naive techniques like replacement of all pixels in the mask with the lowest value, highest value and the average value, and Dimension Reduction significantly outperformed them in the downstream machine learning task, by up to $15\%$ (ImageNet) and is able to match that of Vision Transformers with significantly lower computations. This can be attributed to the fact that Dimension Reduction carefully preserves a certain amount of information in the mask, and is specifically helpful when there are false positives generated upstream. Notably, ODDR provides a performance gain of up to $13\%$ over just Dimension Reduction \cite{defensivedr}, which can be primarily attributed to the Outlier Detection operation.  

\subsection{Impact of ODDR on Model Interpretability}
\begin{figure}[!htbp]
\centerline{\includegraphics[width=0.8\columnwidth]{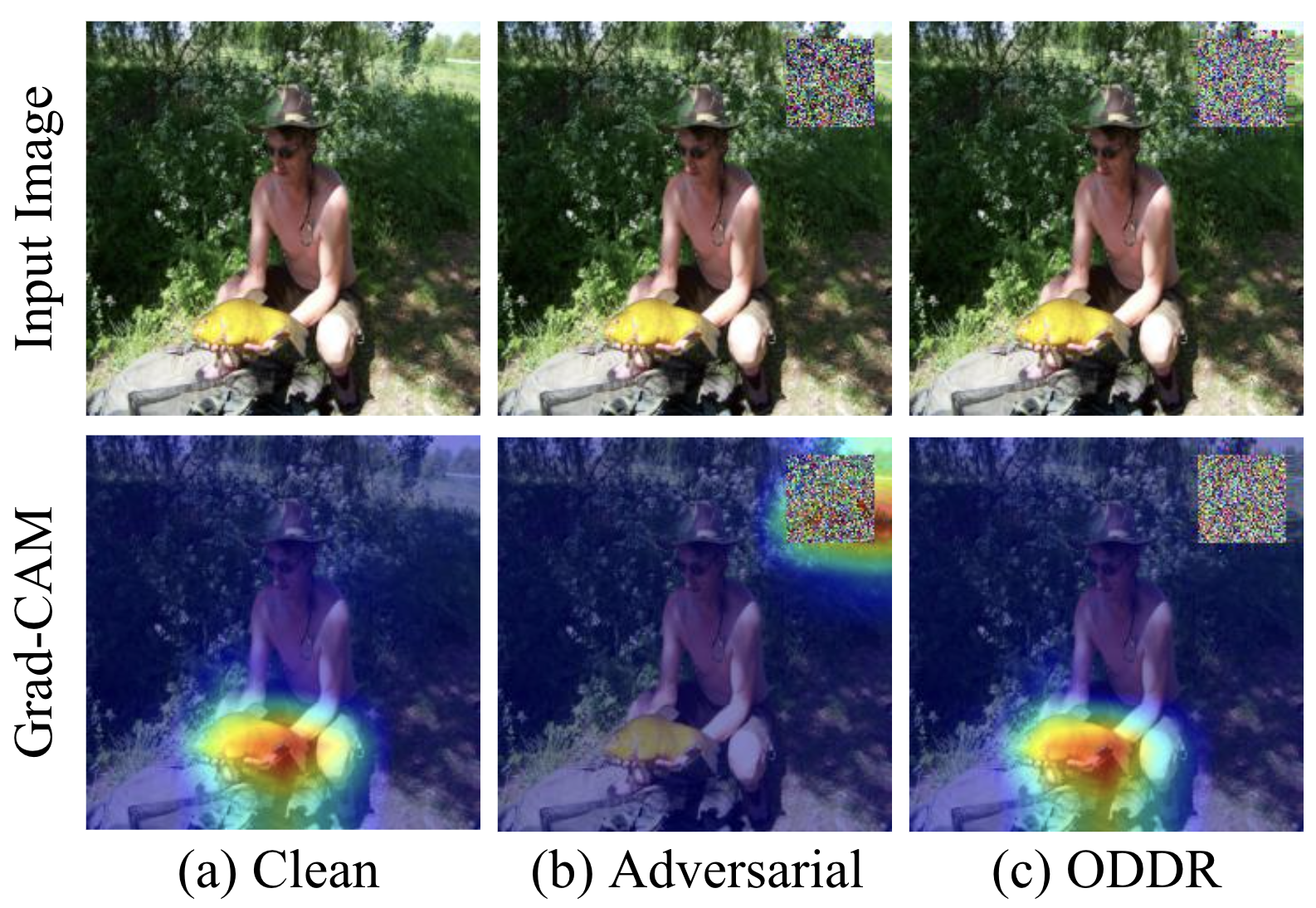}} 
\caption{Grad-CAM visualization result for ODDR in action.} 
\label{fig:grad_cam}
\end{figure}

We conduct a detailed visualization of the Grad-CAM \cite{grad} output for the adversarial example, both before and after the use of our defense. Notably, as shown in Figure \ref{fig:grad_cam}, we observe a significant shift in the network's focus when the adversarial patch is applied, as the attention shifts from the actual object to the manipulated patch. However, upon applying our defense, we observe a restoration of the network's attention back to the original object, highlighting the efficacy of our defense in preserving the model interpretability.

\subsection{Computational Cost}
Our defense mechanism, ODDR, is designed with low computational overhead, making it suitable for deployment on edge devices without requiring GPUs. The computational complexity of the Isolation Forest is of the order $O(n*t*\log s)$ where $s$ is the sub-sample size, $n$ is the number of test samples, and $t$ is the number of trees. The Truncated SVD has a complexity of $O(d^3)$  where $d$ represents the dimensionality, which is typically small as it correlates with the size of the detected outlier clusters.
In practical terms, the average processing time per sample is as follows: 0.034 seconds for ImageNet and CalTech-101 datasets, and 0.041 seconds per frame for the INRIA and CASIA datasets, when executed on an AMD Ryzen 5955 CPU. This is compared to the baseline processing time of 0.019 seconds on NVIDIA GPUs using ResNet50. 
\subsection{Hyper-parameter Tuning}
ODDR uses two categories of hyper-parameters: \textit{Active hyper-parameters} and \textit{Passive hyper-parameters}. The active hyper-parameters significantly influence the performance of ODDR, and the results reported in the tables correspond to their optimal tuned values. These include $c$ (the confidence level for identifying anomalies, ranging from 0.8 to 0.95), $d$ (the size of the mask applied to outlier fragments, between 5 and 50 pixels), and $inf$ (the information preserved after SVD, ranging from 0.7 to 0.9). The passive hyper-parameters, while less impactful on performance, were kept consistent across most experiments. These include $k$ (kernel size, between 20 and 40 pixels), $str$ (stride length, set to $k/2$, $T$ (the number of trees in the Isolation Forest, fixed at 100), the sampling fraction for Isolation Trees (30\%), and the minimum number of fragments required to form an outlier cluster (20-30). Additional details on these parameters can be found in the Appendix.


\subsection{ODDR vs. Naturalistic Patches}
A recent adversarial patch generation method that poses a potentially greater challenge to our defense is the Naturalistic Patch \cite{Hu21, guesmi2023dap}. We investigate the effectiveness of these attack strategies, considering its ability to generate stealthy adversarial patches that mimic real objects, rendering them less visually suspicious and, consequently, more challenging to detect. The results, specifically the recovery rate upon using ODDR, are presented in Tables \ref{tab:detection_Naturalistic}. 
Despite the naturalistic nature of this patch and its seamless integration into the image, our defense successfully mitigated its adversarial impact. As an illustration, our defense demonstrates a notably robust average precision, surpassing that of previously proposed defenses such as LGS  and Jedi. Specifically, our defense achieves a $65\%$ higher average precision compared to LGS, and $55\%$ and $63\%$ for the INRIA dataset, respectively.

\begin{figure}[!htbp]
\centerline{\includegraphics[width=\columnwidth]{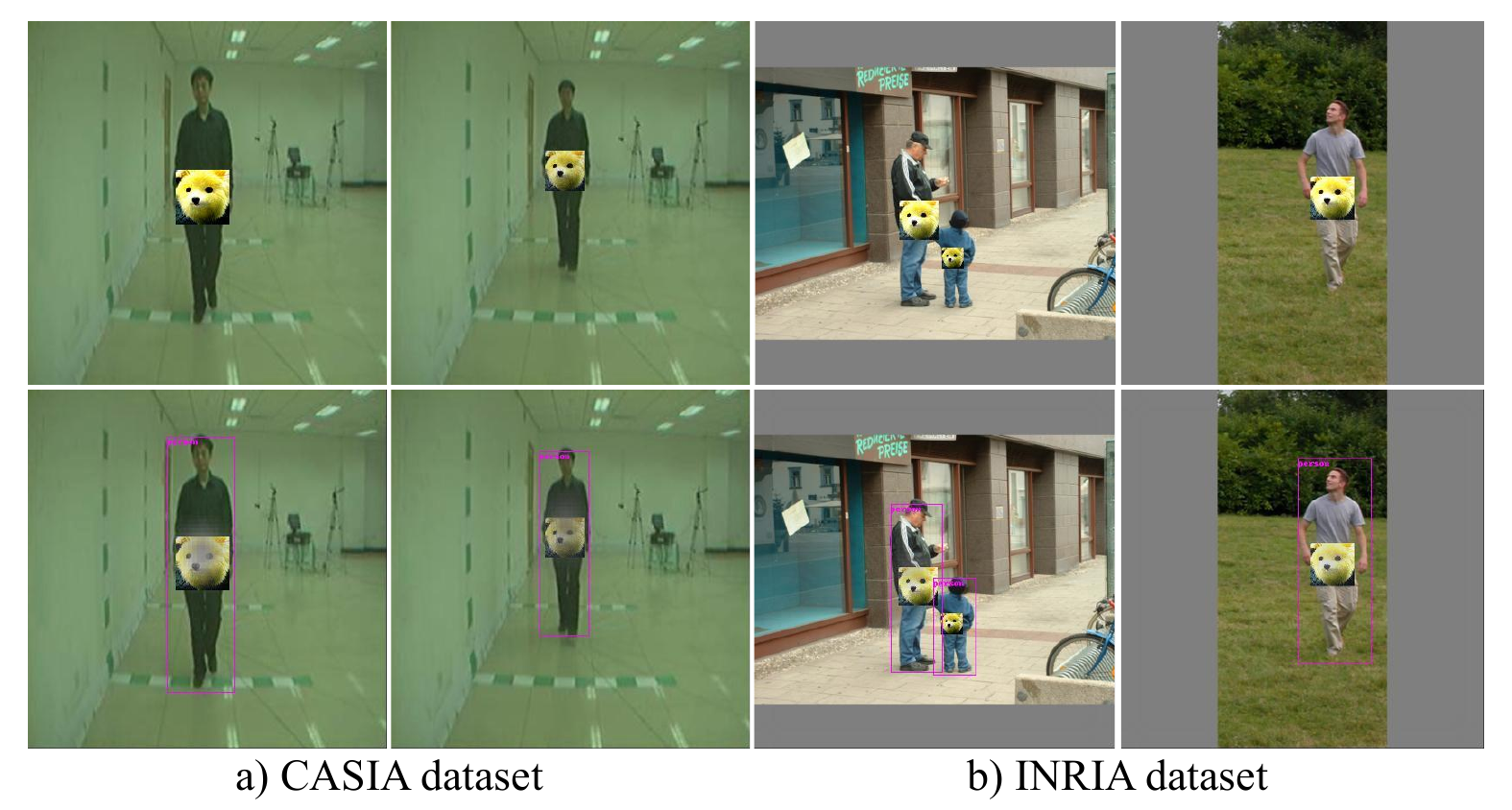}} 
\caption{Illustration of the impact of ODDR on Naturalistic patch-based attacks \mbox{\cite{Hu21}}: \textit{Top row}: Adversarial images. \textit{Bottom row}: Defended images for two datasets: a) CASIA representing indoor settings, and b) INRIA representing outdoor settings.}
\label{fig:_res_nap}
\end{figure}

\begin{table}[!htbp]
    \resizebox{\columnwidth}{!}{%
    \centering
    \begin{tabular}{c|c|c|c|c}
    \hline
             &   \multicolumn{2}{|c|}{INRIA} & \multicolumn{2}{|c}{CASIA} \\
     Defense    & Robust Avg.  & Recovery   & Robust Avg.   & Recovery  \\
                &  Precision  &  Rate   &  Precision   &  Rate      \\
    \hline \hline 
    LGS \cite{naseer2019local}  & 55\%  &  29\% & 50\%  &  48\% \\
    Jedi \cite{jedi}          & 63\%  & \textbf{51\%}   & 67\%  &   76\%  \\
     Ours         (ODDR)            & \textbf{65\%} & 47\%   & \textbf{87.5\%}    & \textbf{90.27\%}   \\
    \hline
    \end{tabular}
    }
    \caption{Performance of our proposed defense compared to two state-of-the-art defenses against Naturalistic Patch \cite{Hu21} attack for INRIA and CASIA dataset.}
    \label{tab:detection_Naturalistic}
\end{table}

\subsection{Adaptive Attack}

Adaptive attacks exploit full knowledge of the defense mechanism to bypass it. In ODDR, this would involve targeting the Isolation Forest-based outlier detection. However, Isolation Forests are non-differentiable, preventing their direct integration into standard attack optimization processes. To circumvent this, we map the adversarial patch's variability to that of randomly selected image fragments across all channels, fitting Gaussian distributions to each fragment. We constrain the patch's updates so its distribution parameters $\mu$ and $\sigma$) align with the averaged fragment parameters ($m$ and $std$), within tolerances $0 \leq |\mu-m|\leq 0.5$  and $1 \leq \sigma/std \leq 2$ (dataset specific). Beyond these ranges, the Gaussian distributions are statistically separable.
On ResNet50 with ImageNet, this approach reduces clean accuracy from 78\% to 73\%, with a robust accuracy of 76\%. Further details are in the Appendix.
\section{Conclusion}
We propose ODDR, specifically designed to counteract the formidable threat posed by patch-based adversarial attacks on machine learning models. Adversarial attacks, particularly those involving localized patches, have emerged as significant challenges, jeopardizing the reliability and security of deep neural networks.
This technique leverages a three-stage pipeline—Fragmentation, Segregation, and Neutralization — which effectively identifies and mitigates adversarial patches while preserving crucial information for machine learning tasks. Through rigorous experimentation on benchmark datasets and state-of-the-art adversarial patches, we substantiate its effectiveness on robustness on multiple machine learning tasks, whilst demonstrating minimal loss in clean accuracies.

{
    \small
    \bibliographystyle{ieeenat_fullname}
    \bibliography{main}

\begin{thebibliography}{39}
\providecommand{\natexlab}[1]{#1}
\providecommand{\url}[1]{\texttt{#1}}
\expandafter\ifx\csname urlstyle\endcsname\relax
  \providecommand{\doi}[1]{doi: #1}\else
  \providecommand{\doi}{doi: \begingroup \urlstyle{rm}\Url}\fi

\bibitem[Athalye et~al.(2018)Athalye, Engstrom, Ilyas, and Kwok]{eot}
Anish Athalye, Logan Engstrom, Andrew Ilyas, and Kevin Kwok.
\newblock Synthesizing robust adversarial examples.
\newblock In \emph{International conference on machine learning}, pages 284--293. PMLR, 2018.

\bibitem[Bochkovskiy et~al.(2020)Bochkovskiy, Wang, and Liao]{yolov4}
Alexey Bochkovskiy, Chien-Yao Wang, and Hong-Yuan~Mark Liao.
\newblock Yolov4: Optimal speed and accuracy of object detection.
\newblock \emph{arXiv preprint arXiv:2004.10934}, 2020.

\bibitem[Brown(2017)]{googleap}
Tom Brown.
\newblock Adversarial patch.
\newblock 2017.

\bibitem[Chattopadhyay et~al.(2019)Chattopadhyay, Chattopadhyay, Gupta, and Kasper]{cod_1}
Nandish Chattopadhyay, Anupam Chattopadhyay, Sourav~Sen Gupta, and Michael Kasper.
\newblock Curse of dimensionality in adversarial examples.
\newblock In \emph{2019 International Joint Conference on Neural Networks (IJCNN)}, pages 1--8. IEEE, 2019.

\bibitem[Chattopadhyay et~al.(2021)Chattopadhyay, Chatterjee, and Chattopadhyay]{cod_2}
Nandish Chattopadhyay, Subhrojyoti Chatterjee, and Anupam Chattopadhyay.
\newblock Robustness against adversarial attacks using dimensionality.
\newblock In \emph{International Conference on Security, Privacy, and Applied Cryptography Engineering}, pages 226--241. Springer, 2021.

\bibitem[Chattopadhyay et~al.(2023)Chattopadhyay, Guesmi, Hanif, Ouni, and Shafique]{defensivedr}
Nandish Chattopadhyay, Amira Guesmi, Muhammad~Abdullah Hanif, Bassem Ouni, and Muhammad Shafique.
\newblock Defensivedr: Defending against adversarial patches using dimensionality reduction.
\newblock \emph{arXiv preprint arXiv:2311.12211}, 2023.

\bibitem[Chawla et~al.(2023)Chawla, Jeeveswaran, Arani, and Zonooz]{mimdepth}
Hemang Chawla, Kishaan Jeeveswaran, Elahe Arani, and Bahram Zonooz.
\newblock Image masking for robust self-supervised monocular depth estimation, 2023.

\bibitem[Chen et~al.(2022)Chen, Li, Wu, Xu, Ding, and Zhang]{chen2022shape}
Zhaoyu Chen, Bo Li, Shuang Wu, Jianghe Xu, Shouhong Ding, and Wenqiang Zhang.
\newblock Shape matters: deformable patch attack.
\newblock In \emph{European conference on computer vision}, pages 529--548. Springer, 2022.

\bibitem[Chen et~al.(2023)Chen, Dash, and Pattabiraman]{Jujutsu}
Zitao Chen, Pritam Dash, and Karthik Pattabiraman.
\newblock Jujutsu: A two-stage defense against adversarial patch attacks on deep neural networks.
\newblock In \emph{Proceedings of the 2023 ACM Asia Conference on Computer and Communications Security}, page 689–703, New York, NY, USA, 2023. Association for Computing Machinery.

\bibitem[Dalal and Triggs(2005)]{inria}
N. Dalal and B. Triggs.
\newblock Histograms of oriented gradients for human detection.
\newblock In \emph{2005 IEEE Computer Society Conference on Computer Vision and Pattern Recognition (CVPR'05)}, pages 886--893 vol. 1, 2005.

\bibitem[De~Maesschalck et~al.(2000)De~Maesschalck, Jouan-Rimbaud, and Massart]{mahal}
Roy De~Maesschalck, Delphine Jouan-Rimbaud, and D{\'e}sir{\'e}~L Massart.
\newblock The mahalanobis distance.
\newblock \emph{Chemometrics and intelligent laboratory systems}, 50\penalty0 (1):\penalty0 1--18, 2000.

\bibitem[Deng et~al.(2009)Deng, Dong, Socher, Li, Li, and Fei-Fei]{imagenet}
Jia Deng, Wei Dong, Richard Socher, Li-Jia Li, Kai Li, and Li Fei-Fei.
\newblock Imagenet: A large-scale hierarchical image database.
\newblock In \emph{2009 IEEE Conference on Computer Vision and Pattern Recognition}, pages 248--255, 2009.

\bibitem[Ester et~al.(1996)Ester, Kriegel, Sander, Xu, et~al.]{dbscan}
Martin Ester, Hans-Peter Kriegel, J{\"o}rg Sander, Xiaowei Xu, et~al.
\newblock A density-based algorithm for discovering clusters in large spatial databases with noise.
\newblock In \emph{kdd}, pages 226--231, 1996.

\bibitem[Freedman et~al.(1978)Freedman, Pisani, and Purves]{freedman}
D Freedman, R Pisani, and R Purves.
\newblock Statistics. 2007.
\newblock \emph{ISBN: 0-393970-833}, 1978.

\bibitem[Geiger et~al.(2012)Geiger, Lenz, and Urtasun]{kitti}
Andreas Geiger, Philip Lenz, and Raquel Urtasun.
\newblock Are we ready for autonomous driving? the kitti vision benchmark suite.
\newblock In \emph{2012 IEEE conference on computer vision and pattern recognition}, pages 3354--3361. IEEE, 2012.

\bibitem[Goodfellow et~al.(2015)Goodfellow, Shlens, and Szegedy]{Goodfellow2015ExplainingAH}
I.~J. Goodfellow, J. Shlens, and C. Szegedy.
\newblock Explaining and harnessing adversarial examples.
\newblock \emph{CoRR}, abs/1412.6572, 2015.

\bibitem[Guesmi et~al.(2023)Guesmi, Hanif, Ouni, and Shafique]{guesmi2023physical}
Amira Guesmi, Muhammad~Abdullah Hanif, Bassem Ouni, and Muhammad Shafique.
\newblock Physical adversarial attacks for camera-based smart systems: Current trends, categorization, applications, research challenges, and future outlook.
\newblock \emph{IEEE Access}, 2023.

\bibitem[Guesmi et~al.(2024{\natexlab{a}})Guesmi, Ding, Hanif, Alouani, and Shafique]{guesmi2023dap}
Amira Guesmi, Ruitian Ding, Muhammad~Abdullah Hanif, Ihsen Alouani, and Muhammad Shafique.
\newblock Dap: A dynamic adversarial patch for evading person detectors.
\newblock In \emph{Proceedings of the IEEE/CVF Conference on Computer Vision and Pattern Recognition}, pages 24595--24604, 2024{\natexlab{a}}.

\bibitem[Guesmi et~al.(2024{\natexlab{b}})Guesmi, Hanif, Alouani, Ouni, and Shafique]{guesmi2024ssap}
Amira Guesmi, Muhammad~Abdullah Hanif, Ihsen Alouani, Bassem Ouni, and Muhammad Shafique.
\newblock Ssap: A shape-sensitive adversarial patch for comprehensive disruption of monocular depth estimation in autonomous navigation applications.
\newblock \emph{arXiv preprint arXiv:2403.11515}, 2024{\natexlab{b}}.

\bibitem[Hariri et~al.(2019)Hariri, Kind, and Brunner]{IFextended}
Sahand Hariri, Matias~Carrasco Kind, and Robert~J Brunner.
\newblock Extended isolation forest.
\newblock \emph{IEEE transactions on knowledge and data engineering}, 33\penalty0 (4):\penalty0 1479--1489, 2019.

\bibitem[He et~al.(2016)He, Zhang, Ren, and Sun]{he2016deep}
Kaiming He, Xiangyu Zhang, Shaoqing Ren, and Jian Sun.
\newblock Deep residual learning for image recognition.
\newblock In \emph{Proceedings of the IEEE conference on computer vision and pattern recognition}, pages 770--778, 2016.

\bibitem[Hu et~al.(2021)Hu, Chen, Kung, Hua, and Tan]{Hu21}
Yu-Chih-Tuan Hu, Jun-Cheng Chen, Bo-Han Kung, Kai-Lung Hua, and Daniel~Stanley Tan.
\newblock Naturalistic physical adversarial patch for object detectors.
\newblock In \emph{2021 IEEE/CVF International Conference on Computer Vision (ICCV)}, pages 7828--7837, 2021.

\bibitem[Karmon(2018)]{lavan}
Dan Karmon.
\newblock Lavan: Localized and visible adversarial noise.
\newblock In \emph{International Conference on Machine Learning}, 2018.

\bibitem[Levine and Feizi(2020)]{levine2020randomized}
Alexander Levine and Soheil Feizi.
\newblock (de) randomized smoothing for certifiable defense against patch attacks.
\newblock \emph{Advances in Neural Information Processing Systems}, 33:\penalty0 6465--6475, 2020.

\bibitem[Li et~al.(2022)Li, Andreeto, Ranzato, and Perona]{caltech}
Fei-Fei Li, Marco Andreeto, Marc'Aurelio Ranzato, and Pietro Perona.
\newblock Caltech 101, 2022.

\bibitem[Li and Ji(2021)]{li2021generative}
Xiang Li and Shihao Ji.
\newblock Generative dynamic patch attack.
\newblock \emph{arXiv preprint arXiv:2111.04266}, 2021.

\bibitem[Liu et~al.(2008)Liu, Ting, and Zhou]{IF}
Fei~Tony Liu, Kai~Ming Ting, and Zhi-Hua Zhou.
\newblock Isolation forest.
\newblock In \emph{2008 eighth ieee international conference on data mining}, pages 413--422. IEEE, 2008.

\bibitem[Liu et~al.(2021)Liu, Lin, Cao, Hu, Wei, Zhang, Lin, and Guo]{swin}
Ze Liu, Yutong Lin, Yue Cao, Han Hu, Yixuan Wei, Zheng Zhang, Stephen Lin, and Baining Guo.
\newblock Swin transformer: Hierarchical vision transformer using shifted windows.
\newblock In \emph{Proceedings of the IEEE/CVF international conference on computer vision}, pages 10012--10022, 2021.

\bibitem[McLachlan(1999)]{mahal_2}
Goeffrey~J McLachlan.
\newblock Mahalanobis distance.
\newblock \emph{Resonance}, 4\penalty0 (6):\penalty0 20--26, 1999.

\bibitem[Naseer et~al.(2019)Naseer, Khan, and Porikli]{naseer2019local}
Muzammal Naseer, Salman Khan, and Fatih Porikli.
\newblock Local gradients smoothing: Defense against localized adversarial attacks.
\newblock In \emph{2019 IEEE Winter Conference on Applications of Computer Vision (WACV)}, pages 1300--1307. IEEE, 2019.

\bibitem[Penny(1996)]{mahal_1}
Kay~I Penny.
\newblock Appropriate critical values when testing for a single multivariate outlier by using the mahalanobis distance.
\newblock \emph{Journal of the Royal Statistical Society: Series C (Applied Statistics)}, 45\penalty0 (1):\penalty0 73--81, 1996.

\bibitem[Simonyan and Zisserman(2014)]{simonyan2014very}
Karen Simonyan and Andrew Zisserman.
\newblock Very deep convolutional networks for large-scale image recognition.
\newblock \emph{arXiv preprint arXiv:1409.1556}, 2014.

\bibitem[Subramanya et~al.(2018)Subramanya, Pillai, and Pirsiavash]{grad}
Akshayvarun Subramanya, Vipin Pillai, and Hamed Pirsiavash.
\newblock Towards hiding adversarial examples from network interpretation.
\newblock \emph{CoRR}, abs/1812.02843, 2018.

\bibitem[Tarchoun et~al.(2023)Tarchoun, Ben~Khalifa, Mahjoub, Abu-Ghazaleh, and Alouani]{jedi}
Bilel Tarchoun, Anouar Ben~Khalifa, Mohamed~Ali Mahjoub, Nael Abu-Ghazaleh, and Ihsen Alouani.
\newblock Jedi: Entropy-based localization and removal of adversarial patches.
\newblock In \emph{Proceedings of the IEEE/CVF Conference on Computer Vision and Pattern Recognition}, pages 4087--4095, 2023.

\bibitem[Thys et~al.(2019)Thys, Ranst, and Goedem{\'{e}}]{thys2019}
Simen Thys, Wiebe~Van Ranst, and Toon Goedem{\'{e}}.
\newblock Fooling automated surveillance cameras: adversarial patches to attack person detection.
\newblock \emph{CoRR}, abs/1904.08653, 2019.

\bibitem[Xia et~al.(2017)Xia, Xu, and Nan]{inception}
Xiaoling Xia, Cui Xu, and Bing Nan.
\newblock Inception-v3 for flower classification.
\newblock In \emph{2017 2nd international conference on image, vision and computing (ICIVC)}, pages 783--787. IEEE, 2017.

\bibitem[Xiang et~al.(2021)Xiang, Bhagoji, Sehwag, and Mittal]{xiang2021patchguard}
Chong Xiang, Arjun~Nitin Bhagoji, Vikash Sehwag, and Prateek Mittal.
\newblock $\{$PatchGuard$\}$: A provably robust defense against adversarial patches via small receptive fields and masking.
\newblock In \emph{30th USENIX Security Symposium (USENIX Security 21)}, pages 2237--2254, 2021.

\bibitem[Yu et~al.(2006)Yu, Tan, and Tan]{casia}
Shiqi Yu, Daoliang Tan, and Tieniu Tan.
\newblock A framework for evaluating the effect of view angle, clothing and carrying condition on gait recognition.
\newblock In \emph{18th international conference on pattern recognition (ICPR'06)}, pages 441--444. IEEE, 2006.

\bibitem[Yuan et~al.(2021)Yuan, Chen, Wang, Yu, Shi, Jiang, Tay, Feng, and Yan]{VIT}
Li Yuan, Yunpeng Chen, Tao Wang, Weihao Yu, Yujun Shi, Zi-Hang Jiang, Francis~EH Tay, Jiashi Feng, and Shuicheng Yan.
\newblock Tokens-to-token vit: Training vision transformers from scratch on imagenet.
\newblock In \emph{Proceedings of the IEEE/CVF international conference on computer vision}, pages 558--567, 2021.

\end{thebibliography}
}

\section{Appendices}
\label{supp}

\subsection{Defense Mechanism}
\label{algorithms}
\begin{figure}[!htp]
\centerline{\includegraphics[width=\columnwidth]{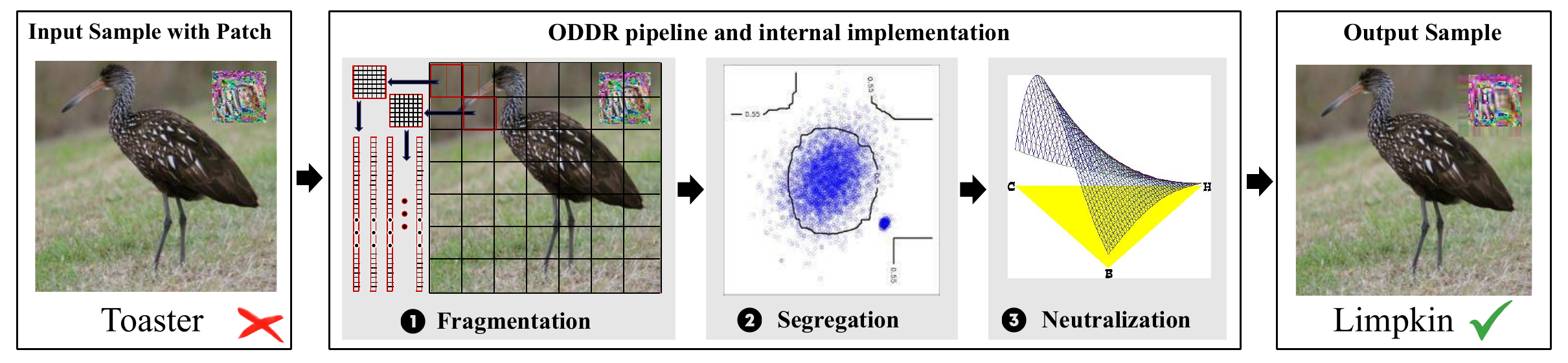}} 
\caption{Overview of the Proposed ODDR Defense Methodology: The three-stage pipeline—Fragmentation, Segregation, and Neutralization—demonstrating the process of identifying and mitigating adversarial patches in the input image features. }
\label{fig:concept_1}
\end{figure}

To briefly reiterate the proposed ODDR mechanism, we make use of a three step process, which helps in outlier detection amongst the fragments and then we neutralize the same using dimension reduction. This is explained in Figure \ref{fig:concept_1}. 

The primary algorithm, which is the ODDR algorithm is presented in the main paper. The Isolation Forest (Algorithm \ref{forest}) makes use of the Isolation Trees which are generated as explained in Algorithm \ref{tree}. These are mentioned hereafter in detail. 
Isolation Forest exhibits swift convergence even with a minimal number of trees, and the incorporation of sub-sampling enhances its ability to yield favorable results while maintaining computational efficiency. The algorithm operates through a recursive process of generating partitions on the dataset, achieved by randomly selecting features and subsequently assigning random split values for these features. An intriguing facet arises from the hypothesis that anomalies necessitate fewer random partitions for isolation compared to their "normal" counterparts in the dataset. Consequently, this characteristic enables Isolation Forest to efficiently and effectively discern anomalies, solidifying its position as a powerful tool for anomaly detection, particularly in datasets with high dimensionality.

\begin{algorithm}
\caption{$IsolationForest(X,T,size)$}
\label{forest}
$\boldsymbol{IN}$: $X = (x_1, \ldots, x_n)$: input data, $T$: number of trees in the forest, $s$: sample size\\
$\boldsymbol{OUT}$: an $IsolationForest$ ($IsolationTree$s ensemble)\\
$\textbf{initialize}: IsolationForest$ \\
$\textbf{set}$ $l_{max} = ceiling(log_{2}$ $ s)$ \\
$\textbf{for}$ $i=1$ to $T$ do: \\
$X' \gets Sample(X,s)$ \\
$IsolationForest \gets IsolationForest \cup IsolationTree(X',0,l_{max})$ (From Algo \ref{tree})\\
$\textbf{end}$ $\textbf{for}$ \\
$\textbf{return}$ $IsolationForest$
\end{algorithm}

\begin{algorithm}
\caption{$IsolationTree(X,h,l_{max})$}
\label{tree}
$\boldsymbol{IN}$: $X = (x_1, \ldots, x_n)$: input data, $h$: height of tree, $l_{max}$: maximum height of tree, $Q$: attributes of $X$ \\
$\boldsymbol{OUT}$: an $IsolationTree$ \\
$\textbf{if}$ $h \geq l_{max}$ or $|X| \leq 1$, then: \\
$\textbf{return}$ $externalNode\{Size \gets |X|\}$ \\
$\textbf{else}$ \\
$X_{left}$ $\gets$ $Select(X,q < p)$, where $min(q) \leq p < max(q)$, $q \in Q$ \\
$X_{right}$ $\gets$ $Select(X,q \geq p)$, where $min(q) \leq p < max(q)$, $q \in Q$ \\
$\boldsymbol{return}$ $internalNode$ \\
\{ \\
$LeftTree$ $\gets$ $IsolationTree(X_{left}, h+1, l_{max})$ \\
$RightTree$ $\gets$ $IsolationTree(X_{right}, h+1, l_{max})$ \\
$attribute$ $\gets$ $q$ \\
$value$ $\gets$ $p$ \\
\} \\
$\boldsymbol{end}$ $\boldsymbol{if}$ \\
\end{algorithm}
\subsection{Patch-based Attacks}

In this evaluation, we employ seven state-of-the-art adversarial patches to rigorously assess model performance. For classification tasks, we utilize the Adversarial Patch (GAP) \cite{googleap}, LAVAN \cite{lavan}, Generative Dynamic Patch Attack (GDPA) \cite{li2021generative}, and Shape Matters (SM) \cite{chen2022shape}. 

\textit{Adversarial Patch (GAP)} \cite{googleap} offers a more practical form of attack for real-world scenarios compared to Lp-norm-based adversarial perturbations, which require object capture through a camera. This attack creates universal patches that can be applied anywhere. Additionally, the attack incorporates Expectation over Transformation (EOT) \cite{eot} to enhance the strength of the generated adversarial patch.

\textit{LAVAN} \cite{lavan} is a technique for generating localized and visible patches that can be applied across various images and locations. This approach involves training the patch iteratively by selecting a random image and placing it at a randomly chosen location. This iterative process makes sure that the model can capture the distinguishing features of the patch across a range of scenarios, thereby enhancing its ability to transfer and its overall effectiveness.

\textit{Generative Dynamic Patch Attack (GDPA}) \cite{li2021generative}  is a method that adversarially generates both the patch pattern and its location for each input image. It is presented as a versatile attack framework capable of producing dynamic or static, as well as visible or invisible, patches with minimal configuration adjustments. By utilizing a generator to create both the patch pattern and its location simultaneously for each image, GDPA significantly reduces inference time. Additionally, GDPA can be easily incorporated into adversarial training to enhance a model's robustness against various adversarial attacks.

Shape Matters (SM) \cite{chen2022shape} proposes a \textit{Deformable Patch Representation (DPR)}, which leverages the geometric structure of triangles to enable a differentiable mapping between contour modeling and masks, allowing the shape to be deformably adjusted during patch generation. Building on DPR, a shape and texture joint optimization algorithm for adversarial patches, termed DAPatch, is introduced. This algorithm effectively optimizes both shape and texture to enhance attack performance. DPR also explicitly examines the importance of shape information on the robustness of deep neural networks (DNNs) from an adversarial perspective, contributing to a deeper understanding of the inherent vulnerabilities of DNNs.

For detection tasks, we use the \textit{AdvYOLO} adversarial patch as introduced in \cite{thys2019}. This method generates a small, compact patch that, when held by an attacker, can effectively deceive the YOLO detector \cite{yolov4}. Typically, deep neural network-based detectors are designed to predict bounding box positions, object probabilities, and class scores for objects within an input image. However, AdvYOLO modifies this process by employing a training approach that minimizes the object probabilities and class scores for the target class—specifically, people. As a result, the detector is tricked into ignoring the presence of individuals, rendering them undetected.

The \textit{Naturalistic Patch} \cite{Hu21} leverages the efficiency of generative adversarial networks (GANs) in crafting physical adversarial patches designed to deceive person detection systems. In their novel approach, Hu et al. \cite{Hu21} focus on manipulating the detection probabilities for the person class. By applying the principles of AdvYOLO, they minimize the objectness and class probabilities specifically targeting the person class, thus creating patches that effectively evade person detectors.

For monocular depth estimation (MDE) models, we employ the \textit{shape-sensitive adversarial patch (SSAP)} as described in \cite{guesmi2024ssap}. SSAP is specifically designed to disrupt monocular depth estimation systems used in autonomous navigation. By introducing an adaptive adversarial patch, SSAP seeks to mislead depth estimation models, resulting in incorrect depth maps, which can critically affect the performance of autonomous systems.
A key characteristic of SSAP is its adaptability to different environments. The patch is optimized through an iterative process that updates the patch using gradients derived from the depth estimation model's backpropagation. This iterative optimization allows SSAP to dynamically adjust and maintain its effectiveness across various scenes, thereby maximizing its ability to deceive the depth estimation model.

\subsection{Evaluation Metrics}

\subsubsection{Classification Task:}
The metric that has been used for the Classification Task is Top-1 percentage accuracy on a test sample of 1000 images. Top-1 percentage accuracy here refers to the proportion of test samples where the model's highest-confidence prediction (i.e., the class with the highest probability score) matches the correct label. In other words, it's the percentage of cases where the model's first prediction is accurate.

\subsubsection{Object Detection Task:}
To assess the effectiveness of our defense in object detection tasks, we employ the following metrics: \textit{Robust Average Precision} evaluates the model’s precision averaged across all recall levels for a given class, particularly under adversarial conditions. This metric quantifies the model’s ability to maintain high performance when subjected to attacks, with the defense mechanism in place.
\textit{Recovery Rate} quantifies the proportion of correctly restored outputs by the defense mechanism relative to the total number of successful adversarial attacks. This metric captures the inherent positive impact of the defense by measuring its effectiveness in mitigating the adversarial effects.

\subsubsection{Monocular Depth Estimation Task:}
To evaluate the effectiveness of our proposed defense, we utilize the same metrics as those used in \cite{guesmi2024ssap}: the \textit{mean depth estimation error} ($E_d$) associated with the target object, and the \textit{ratio of the affected region} ($R_a$). For the computation of these metrics, the depth prediction of the clean target object is considered the ground truth.

The mean depth estimation error quantifies the extent to which our proposed adversarial patch impacts the accuracy of depth estimation. A higher value in this metric suggests a more successful attack. Similarly, the ratio of the affected region indicates the extent of the attack's influence, with higher values signifying a more impactful attack.

The \textit{mean depth estimation error} is calculated as follows:
\begin{equation}
    E_{d} = \frac{\sum_{i,j}(\arrowvert d - d_{adv} \arrowvert \odot M_f) }{\sum_{i,j}{M_f }}
\end{equation}
where $d$ represents the clean predicted depth, and $d_{adv}$ is the depth prediction after the adversarial attack.

The \textit{ratio of the affected region} measures the percentage of pixels whose depth values are altered beyond a specific threshold, in relation to the total number of pixels within the focus mask ($M_f$). Pixels with depth changes greater than $0.1$ are considered affected. This metric is calculated as follows:

\begin{equation}
    R_{a} = \frac{\sum_{i,j}I((\arrowvert d - d_{adv} \arrowvert \odot M_f) > 0.1) }{\sum_{i,j}{M_f }}
\end{equation}

Additionally, we use the Mean Square Error (MSE) to evaluate the model's performance concerning the predicted depth map from an unperturbed input. The MSE is computed as:
\begin{equation}
    MSE = \frac{1}{N}\sum_{i,j}(d_{adv_{i,j}} - d_{i,j})^2  
\end{equation}

where $N$ represents the total number of pixels.

\subsection{Additional Results for Classification Task}
\label{classification_results}

Here, we present findings demonstrating the efficacy of our defense mechanism against two newly introduced attack methodologies, across an expanded array of model architectures, including other CNNs and vision transformer models. Additionally, we explore the impact of different hyper-parameters on defense performance.
Table \ref{tab1} presents ODDR's performance against two attacks: Generative Dynamic Patch Attack (GDPA) \cite{li2021generative} \& Shape Matters (SM) \cite{chen2022shape}. 
For Detection tasks, apart from YOLO, we further tested the Faster-RCNN model on a subset of INRIA dataset, which generated a clean accuracy of 100\%, adversarial accuracy of 33.33\% and post ODDR robust accuracy of 86.66\%.

As mentioned in the Results section, our evaluation was focused on measuring the model's robust accuracy. To illustrate the impact of our defense strategy, we initially generated adversarial patches using two distinct attack strategies, namely LAVAN, GoogleAp and GDPA. Subsequently, we reported the model's robust accuracy across different patch sizes and various models, for LAVAN and GoogleAP here, due to restrictions of content volume, the complete set of results including all the patch sizes.
It may also be noted here that the performances of ODDR on clean samples without the patches are $79.8\%$ (ResNet-152), $77.3\%$ (ResNet-50) and $73.2\%$ (VGG-19) on ImageNet and $92.8\%$ (ResNet-152), $88.9\%$ (ResNet-50) and $87.1\%$ (VGG-19) on CalTech-101.

Tables \ref{tab1_imagenet_sup} and \ref{tab2_imagenet_sup} highlight the fact that our proposed ODDR is able to maintain impressive performance across different degrees of potency of attacks, showcasing a remarkable level of robust accuracy when countering GoogleAp and LAVAN attacks on the ImageNet dataset. For instance, our defense achieves robust accuracy rates of $79.1\%$ and $74.1\%$ when employed against GoogleAp and LAVAN attacks, respectively for a smaller patch size of $38$x$38$ pixels and $80.3\%$ and $75.4\%$ on the largest of patch sizes of $50$x$50$ pixels.

\begin{table}[!htbp]
\centering
\resizebox{\columnwidth}{!}{%
\begin{tabular}{c|l|c|c|c}
\hline
\multirow{2}{*}{\begin{tabular}[c]{@{}c@{}}Patch\\ Size\end{tabular}} & \multicolumn{1}{c|}{\multirow{2}{*}{\begin{tabular}[c]{@{}c@{}}Model /\\ Neural Network\end{tabular}}} & \multirow{2}{*}{\begin{tabular}[c]{@{}c@{}}Baseline\\ Accuracy\end{tabular}} & \multirow{2}{*}{\begin{tabular}[c]{@{}c@{}}Adversarial\\ Accuracy\end{tabular}} & \multirow{2}{*}{\begin{tabular}[c]{@{}c@{}}Robustness\\ (w/ patch)\end{tabular}} \\
 & \multicolumn{1}{c|}{} &  &  &  \\ \hline 
\multirow{3}{*}{\begin{tabular}[c]{@{}c@{}}38\\ x\\ 38\end{tabular}} & ResNet 152 & 81.2\% & 39.9\%  & 79.1\%  \\ 
 & ResNet 50 & 78.4\%  & 38.8\%  & 75.6\%  \\ 
 & VGG 19 & 74.2\%  & 39.1\%  & 72.8\%  \\ \hline
\multirow{3}{*}{\begin{tabular}[c]{@{}c@{}}41\\ x\\ 41\end{tabular}} & ResNet 152 & 81.2\% & 21.4\% & 79.6\% \\ 
 & ResNet 50 & 78.4\% & 21.1\% & 77.1\% \\ 
 & VGG 19 & 74.2\% & 22.8\% & 71.9\% \\ \hline
\multirow{3}{*}{\begin{tabular}[c]{@{}c@{}}44\\ x\\ 44\end{tabular}} & ResNet 152 & 81.2\%  & 14.6\%  & 80.1\%  \\ 
 & ResNet 50 & 78.4\%  & 14.2\%  & 76.4\%  \\ 
 & VGG 19 & 74.2\%  & 15.8\%  & 72.1\%  \\ \hline
\multirow{3}{*}{\begin{tabular}[c]{@{}c@{}}47\\ x\\ 47\end{tabular}} & ResNet 152 & 81.2\% & 9.3\% & 78.9\% \\ 
 & ResNet 50 & 78.4\% & 9.0\% & 77.2\% \\ 
 & VGG 19 & 74.2\% & 10.6\% & 72.3\% \\ \hline
\multirow{3}{*}{\begin{tabular}[c]{@{}c@{}}50\\ x\\ 50\end{tabular}} & ResNet 152 & 81.2\%  & 4.9\%  & 80.3\%  \\ 
 & ResNet 50 & 78.4\%  & 4.5\%  & 77.8\%  \\ 
 & VGG 19 & 74.2\%  & 3.8\%  & 72.5\%  \\ \hline
\end{tabular}
}
\caption{ODDR robustness on GoogleAp attack (ImageNet dataset)}
\label{tab1_imagenet_sup}
\end{table}

\begin{table}[!htbp]
\centering
\resizebox{\columnwidth}{!}{%
\begin{tabular}{c|l|c|c|c}
\hline
\multirow{2}{*}{\begin{tabular}[c]{@{}c@{}}Patch\\ Size\end{tabular}} & \multicolumn{1}{c|}{\multirow{2}{*}{\begin{tabular}[c]{@{}c@{}}Model /\\ Neural Network\end{tabular}}} & \multirow{2}{*}{\begin{tabular}[c]{@{}c@{}}Baseline\\ Accuracy\end{tabular}} & \multirow{2}{*}{\begin{tabular}[c]{@{}c@{}}Adversarial\\ Accuracy\end{tabular}} & \multirow{2}{*}{\begin{tabular}[c]{@{}c@{}}Robustness\\ (w/ patch)\end{tabular}} \\
 & \multicolumn{1}{c|}{} &  &  &  \\ \hline 
\multirow{3}{*}{\begin{tabular}[c]{@{}c@{}}38\\ x\\ 38\end{tabular}} & ResNet 152 & 81.2\%  & 10.1\%  & 74.1\%  \\ 
 & ResNet 50 & 78.4\%  & 10.2\%  & 70.2\%  \\ 
 & VGG 19 & 74.2\%  & 11.1\%  & 71.1\%  \\ \hline
\multirow{3}{*}{\begin{tabular}[c]{@{}c@{}}41\\ x\\ 41\end{tabular}} & ResNet 152 & 81.2\% & 7.9\% & 76.9\% \\ 
 & ResNet 50 & 78.4\% & 8.3\% & 74.7\% \\ 
 & VGG 19 & 74.2\% & 8.1\% & 70.1\% \\ \hline
\multirow{3}{*}{\begin{tabular}[c]{@{}c@{}}44\\ x\\ 44\end{tabular}} & ResNet 152 & 81.2\%  & 4.9\%  & 76.3\%  \\ 
 & ResNet 50 & 78.4\%  & 4.8\%  & 72.8\%  \\ 
 & VGG 19 & 74.2\%  & 4.8\%  & 73.1\%  \\ \hline
\multirow{3}{*}{\begin{tabular}[c]{@{}c@{}}47\\ x\\ 47\end{tabular}} & ResNet 152 & 81.2\% & 1.2\% & 78.4\% \\ 
 & ResNet 50 & 78.4\% & 1.0\% & 77\% \\ 
 & VGG 19 & 74.2\% & 1.7\% & 72.1\% \\ \hline
\multirow{3}{*}{\begin{tabular}[c]{@{}c@{}}50\\ x\\ 50\end{tabular}} & ResNet 152 & 81.2\%  & 1.9\%  & 75.4\%  \\ 
 & ResNet 50 & 78.4\%  & 2.0\%  & 74.1\%  \\ 
 & VGG 19 & 74.2\%  & 2.1\%  & 71.8\%  \\ \hline
\end{tabular}
}
\caption{ODDR robustness on LAVAN attack (ImageNet dataset)}
\label{tab2_imagenet_sup}
\end{table}

Exactly as in Section Experimental Results, the extended versions are presented in Tables \ref{tab1_caltech_sup} and \ref{tab2_caltech_sup} for the Caltech-101 dataset. As evident, our proposed ODDR defense technique achieves outstanding performance with robust accuracy rates of $90.8\%$ and $91.1\%$ when defending against GoogleAp and LAVAN attacks, respectively for a smaller patch size of $38$x$38$ pixels and $91.3\%$ and $91.6\%$ on the largest of patch sizes of $50$x$50$.

\begin{table}[!htbp]
\centering
\resizebox{\columnwidth}{!}{%
\begin{tabular}{c|l|c|c|c}
\hline
\multirow{2}{*}{\begin{tabular}[c]{@{}c@{}}Patch\\ Size\end{tabular}} & \multicolumn{1}{c|}{\multirow{2}{*}{\begin{tabular}[c]{@{}c@{}}Model /\\ Neural Network\end{tabular}}} & \multirow{2}{*}{\begin{tabular}[c]{@{}c@{}}Baseline\\ Accuracy\end{tabular}} & \multirow{2}{*}{\begin{tabular}[c]{@{}c@{}}Adversarial\\ Accuracy\end{tabular}} & \multirow{2}{*}{\begin{tabular}[c]{@{}c@{}}Robustness\\ (w/ patch)\end{tabular}} \\
 & \multicolumn{1}{c|}{} &  &  &  \\ \hline 
\multirow{3}{*}{\begin{tabular}[c]{@{}c@{}}38\\ x\\ 38\end{tabular}} & ResNet 152 & 94.1\%  & 48.6\%  & 90.8\%  \\ 
 & ResNet 50 & 90.9\%  & 49.2\%  & 86.4\%  \\ 
 & VGG 19 & 88.6\%  & 47.1\%  & 85.6\%  \\ \hline
\multirow{3}{*}{\begin{tabular}[c]{@{}c@{}}41\\ x\\ 41\end{tabular}} & ResNet 152 & 94.1\% & 30.1\% & 91.2\% \\ 
 & ResNet 50 & 90.9\% & 29.3\% & 87.1\% \\ 
 & VGG 19 & 88.6\% & 28.2\% & 87.3\% \\ \hline
\multirow{3}{*}{\begin{tabular}[c]{@{}c@{}}44\\ x\\ 44\end{tabular}} & ResNet 152 & 94.1\%  & 10.8\%  & 90.2\%  \\ 
 & ResNet 50 & 90.9\%  & 11.2\%  & 86.9\%  \\ 
 & VGG 19 & 88.6\%  & 12.6\%  & 85.4\%  \\ \hline
\multirow{3}{*}{\begin{tabular}[c]{@{}c@{}}47\\ x\\ 47\end{tabular}} & ResNet 152 & 94.1\% & 9.1\% & 91.2\% \\ 
 & ResNet 50 & 90.9\% & 9.6\% & 86.5\% \\ 
 & VGG 19 & 88.6\% & 10.4\% & 86.2\% \\ \hline
\multirow{3}{*}{\begin{tabular}[c]{@{}c@{}}50\\ x\\ 50\end{tabular}} & ResNet 152 & 94.1\%  & 6.8\%  & 91.3\%  \\ 
 & ResNet 50 & 90.9\%  & 5.9\%  & 87.2\%  \\ 
 & VGG 19 & 88.6\%  & 6.2\%  & 85.4\%  \\ \hline
\end{tabular}
}
\caption{ODDR robustness on GoogleAp attack (CalTech-101 dataset)}
\label{tab1_caltech_sup}
\end{table}

\begin{table}[!htbp]
\centering
\resizebox{\columnwidth}{!}{%
\begin{tabular}{c|l|c|c|c}
\hline
\multirow{2}{*}{\begin{tabular}[c]{@{}c@{}}Patch\\ Size\end{tabular}} & \multicolumn{1}{c|}{\multirow{2}{*}{\begin{tabular}[c]{@{}c@{}}Model /\\ Neural Network\end{tabular}}} & \multirow{2}{*}{\begin{tabular}[c]{@{}c@{}}Baseline\\ Accuracy\end{tabular}} & \multirow{2}{*}{\begin{tabular}[c]{@{}c@{}}Adversarial\\ Accuracy\end{tabular}} & \multirow{2}{*}{\begin{tabular}[c]{@{}c@{}}Robustness\\ (w/ patch)\end{tabular}} \\
 & \multicolumn{1}{c|}{} &  &  &  \\ \hline
\multirow{3}{*}{\begin{tabular}[c]{@{}c@{}}38\\ x\\ 38\end{tabular}} & ResNet 152 & 94.1\%  & 15.6\%  & 91.1\%  \\ 
 & ResNet 50 & 90.9\%  & 17.1\%  & 87.3\%  \\ 
 & VGG 19 & 88.6\%  & 15.3\%  & 84.9\%  \\ \hline
\multirow{3}{*}{\begin{tabular}[c]{@{}c@{}}41\\ x\\ 41\end{tabular}} & ResNet 152 & 94.1\% & 14.2\% & 90.9\% \\ 
 & ResNet 50 & 90.9\% & 13.9\% & 86.8\% \\ 
 & VGG 19 & 88.6\% & 13.8\% & 85.8\% \\ \hline
\multirow{3}{*}{\begin{tabular}[c]{@{}c@{}}44\\ x\\ 44\end{tabular}} & ResNet 152 & 94.1\%  & 8.4\%  & 91.3\%  \\ 
 & ResNet 50 & 90.9\%  & 8.9\%  & 87.8\%  \\ 
 & VGG 19 & 88.6\%  & 9.1\%  & 84.8\%  \\ \hline
\multirow{3}{*}{\begin{tabular}[c]{@{}c@{}}47\\ x\\ 47\end{tabular}} & ResNet 152 & 94.1\% & 5.1\% & 90.8\% \\ 
 & ResNet 50 & 90.9\% & 4.9\% & 88.1\% \\ 
 & VGG 19 & 88.6\% & 6.1\% & 86.4\% \\ \hline
\multirow{3}{*}{\begin{tabular}[c]{@{}c@{}}50\\ x\\ 50\end{tabular}} & ResNet 152 & 94.1\%  & 1.2\%  & 91.6\%  \\ 
 & ResNet 50 & 90.9\%  & 1.0\%  & 86.7\%  \\ 
 & VGG 19 & 88.6\%  & 1.8\%  & 85.6\%  \\ \hline
\end{tabular}
}
\caption{ODDR robustness on LAVAN attack (CalTech-101 dataset)}
\label{tab2_caltech_sup}
\end{table}

The overall takeaway from the extensive experimentation is that unlike many other defense techniques available in the literature, the proposed ODDR mechanism is able to successfully thwart the patch based adversarial attacks irrespective of the patch sizes.
Table \ref{tab2_comparison} presents a comparative study of ODDR's robustness against three different attacks across three different neural architectures on the ImageNet dataset. 

\begin{table}[h]
\centering
\caption{ODDR comparison across attacks on ImageNet dataset}
\label{tab2_comparison}
\resizebox{\columnwidth}{!}{%
\begin{tabular}{l|c|cc|cccc}
\hline
\multicolumn{1}{c|}{Model} & Attack & Clean & Adv & LGS & Jedi & DR & ODDR \\ \hline \hline 
\multirow{3}{*}{ResNet 152} & GAP & 81.2 & 39.3 & 53.9 & 64.4 & 66.2 & 79.1 \\
 & LAVAN & 81.2 & 10.1 & 49.9 & 54.6 & 59.4 & 74.1 \\
 & GDPA & 81.2 & 56.4 & 64.2 & 61.5 & 60.2 & 73.1 \\ \hline
\multirow{3}{*}{Inception v3} & GAP & 79.8 & 38.2 & 55.1 & 66.8 & 61.3 & 77.6 \\
 & LAVAN & 79.8 & 10.9 & 50.7 & 59.4 & 55.4 & 77.1 \\
 & GDPA & 79.8 & 62.1 & 66.9 & 71.4 & 58.6 & 72.9 \\ \hline
\multirow{3}{*}{swin\_v2\_b} & GAP & 84.5 & 35.9 & 65.2 & 70.6 & 62.6 & 81.9 \\
 & LAVAN & 84.5 & 6.9 & 60.9 & 57.8 & 53.8 & 81.2 \\
 & GDPA & 84.5 & 69.8 & 76.3 & 68.4 & 58.9 & 81.8 \\ \hline
\end{tabular}
}
\end{table}

\subsection{Hyper-parameter Tuning}
\begin{figure}[!htbp]
\centerline{\includegraphics[width=\columnwidth]{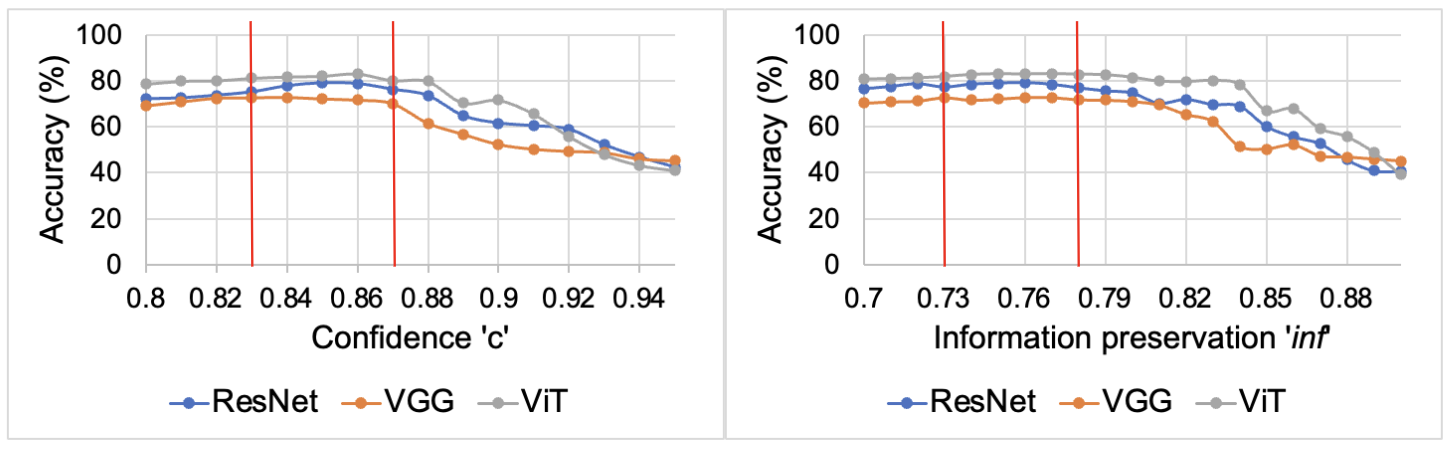}} 
\caption{Hyper-parameter tuning data}
\label{fig:patches}
\end{figure}

As mentioned earlier, ODDR uses two categories of hyper-parameters: \textit{Active hyper-parameters} and \textit{Passive hyper-parameters}. The active hyper-parameters significantly influence the performance of ODDR, and the results reported in the tables correspond to their optimal tuned values. These include $c$ (the confidence level for identifying anomalies, ranging from 0.8 to 0.95) and $inf$ (the information preserved after SVD, ranging from 0.7 to 0.9). Figure \ref{fig:patches} shows the change in the robust accuracy of ODDR with different hyper-parameters for the classification task, on three different neural network models. It is note-worthy that there is a knee-bend in the plots, and selecting any value before that point of bending, within a range, works equally well. 

\subsection{Adaptive Attack details}
In order to generate a patch capable of circumventing our defense, the adversarial noise must adhere to a distribution similar/close to that of clean images.
We implement the adaptive attack by constraining the distribution of the adversarial patch to closely match the average distribution of $n$ randomly selected fragments from the clean image. Specifically, we calculate the average distribution in terms of both mean and standard deviation. We then compute the mean difference and standard deviation ratio between the average distribution of the fragments and the adversarial patch. This helps us in generating an attack which is able to counter the premise of the ODDR defense mechanism. 

\begin{figure}[htbp]
\centerline{\includegraphics[width=\columnwidth]{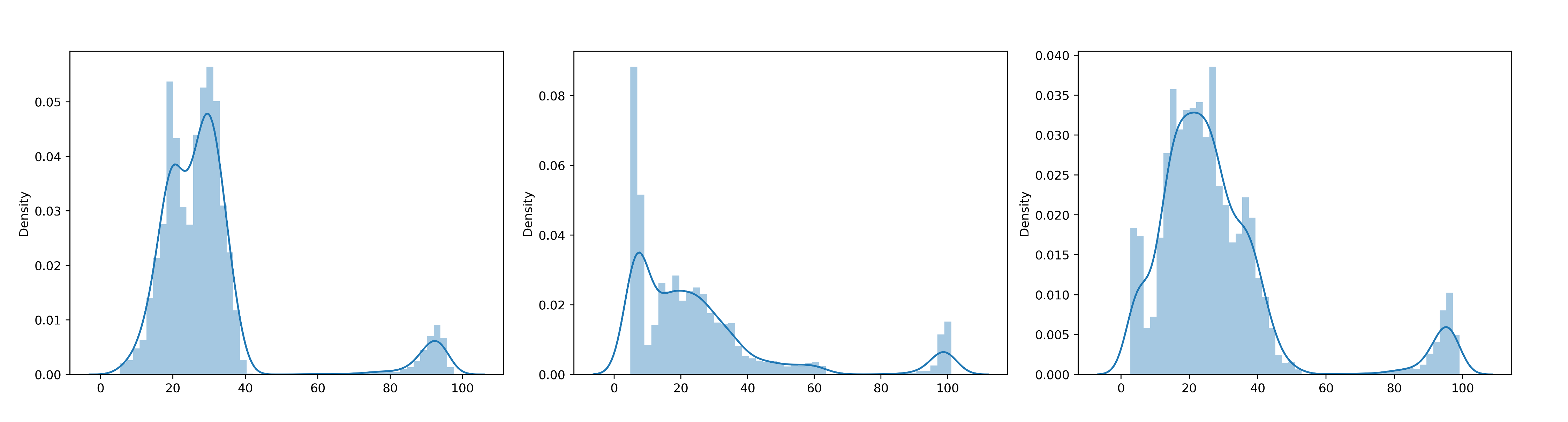}}
\caption{Plot of Mahalanobis distances of fragments to represent anomalous behaviour of adversarial patches as seen in the bi-modal distribution.}
\label{fig:mahal}
\end{figure}

Taking a step back to study whether the adversarial noise contained in the adversarial patches belong to a different distribution or not, as compared to the clean images, we make use of a distance metric to measure the distance between the overall distribution of the image and the adversarial patch. Specifically, we split the adversarial image along with the patch into fragments (as described in Section ODDR Defense Mechanism and fit a distribution to it, and calculate the Mahalanobis distance of every fragment from that distribution \cite{mahal_1}. 
The Mahalanobis distance \cite{mahal} is a distance metric which is designed to measure the distances of data points with respect to a distribution. Formally, for a probability distribution $Q$ on $\mathbb{R}$, which has the mean $\mu = (\mu_1,\ldots, \mu_N)^T$, a positive definite covariance matrix $S$, then for any point $x = (x_1, \ldots, x_N)^T$ for the said distribution $Q$, the Mahalanobis distance is \cite{mahal_2}: 
\begin{equation}
    d_{M} (x,Q) = \sqrt{(x-\mu)^T S^{-1}(x-\mu)} \nonumber
\end{equation}
For our case, the $x_{i}$s are the fragments created as part of the Fragmentation phase. Once we calculate the Mahalanobis distance for every such fragment, and plot them, we observe a bi-modal distribution in Figure \ref{fig:mahal}, with the distance measured for the fragments lying on the adversarial patch having a significantly higher value of Mahalanobis distance than the rest, belonging to a different distribution. This is a clear indication that the patch contains informational variability that is different from the rest of the image and can be isolated as anomalies.  

\begin{figure}[htbp]
\centerline{\includegraphics[width=\columnwidth]{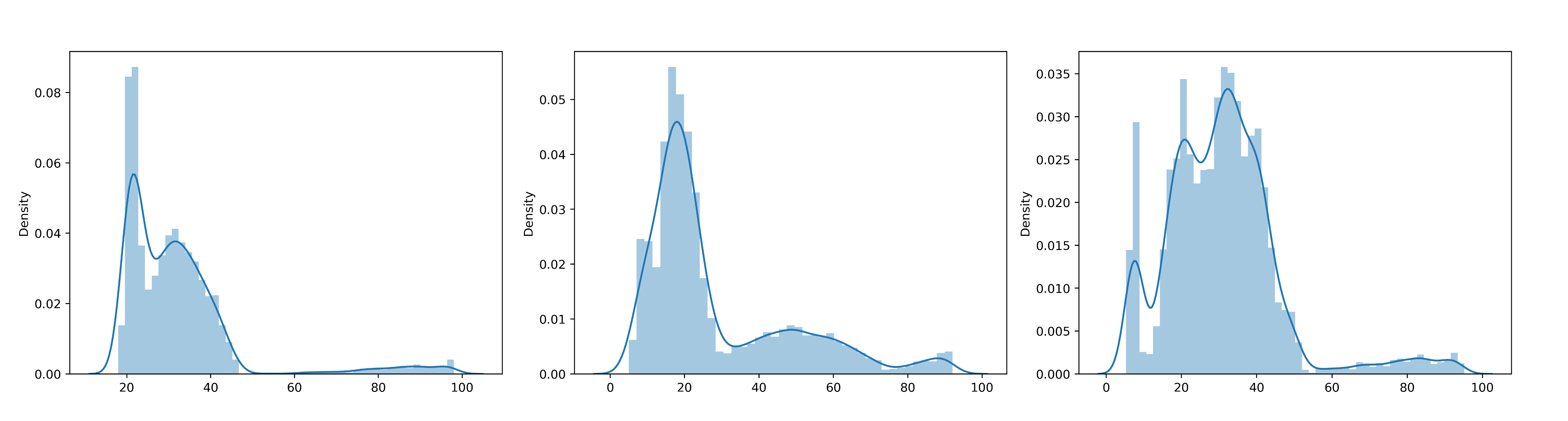}}
\caption{Plot of Mahalanobis distances of fragments upon introducing the adaptive attack, showing one single heavy tailed distribution.}
\label{fig:mahal_ada}
\end{figure}

Once we generated the adversarial patches using the adaptive attack proposed in Section Adaptive Attack in the main paper, we recalculated the Mahalanobis distance for each fragment and plotted them. Upon observation, we noticed a one single heavy tailed distribution indicating that the fragments located on the adversarial patch exhibited a Mahalanobis distance closer to that of the rest of the image fragments (See Figure \ref{fig:mahal_ada}) which makes it challenging to isolate the patch as anomalies.

The key takeaway from this exercise is to showcase the fact that adversarial patches contain information that do not belong to the distribution of the information contained in the rest of the images or video frames. This is precisely why the outlier detection mechanism is successful in isolating the fragments which contain the adversarial patch. If however, we do force the adversarial patch to contain information belonging to the distribution of the rest of the image, by setting constraints while training the patch, as in our proposed adversarial attack in the Adaptive Attack Section , then the patch itself becomes less effective and is unable to bring down the accuracy of the classifier or object detector.

\subsection{Sample Images}
We have tested ODDR for its functionality on various machine learning tasks, datasets and adversarial patches. It is not possible to include all samples as part of the supplementary materials, so a randomly chosen few are added to demonstrate the technique and substantiate the findings. The sample images have been attached in the folder contained within the supplementary materials. The submitted folder should be navigated as mentioned hereafter.  

The root directory is called \textit{SAMPLES}, which contains three folders, for the three types of machine learning applications that we have considered, namely \textit{IMAGE CLASSIFICATION},  \textit{OBJECT DETECTION} and \textit{MONOCULAR DEPTH ESTIMATION}. Within the \textit{CLASSIFICATION} folder, there are two sub-folders, one containing the \textit{ImageNet\_samples} (the naming of the individual files are self-explanatory) and the other containing the samples from \textit{GradCam}, which has been used to understand model interpretability in Section Discussion. Within the \textit{DETECTION} folder, we have two sub-folders corresponding to the two datasets that we have used in the detection task. Each of them, \textit{CASIA} and \textit{INRIA} contain two sub-folders for two different types of adversarial patches that have been used in the experiments in the earlier sections. For each combination of the dataset and adversarial patch, we have the folders containing \textit{samples\_with\_ODDR} (that have the adversarial samples after having undergone the ODDR defence pipeline) and the \textit{inference} which show the efficacy of the ODDR scheme by running the inference model on the ODDR-ed samples. The samples which have undergone the ODDR based defense technique could be tested with the correspondingly appropriate inference model as mentioned in the experiments section for verification. Within the \textit{DEPTH ESTIMATION} folder, there are three sub-folders corresponding to the \textit{Clean Samples}, \textit{Adversarial Samples} where the patch is inserted and the \textit{Defended Samples} which have ODDR implemented on them. The corresponding depth map is also provided for each sample in the respective folders.

\subsection{Code}
This work is supported by a project grant which marks the code as closed IP and proprietary and therefore it is something that cannot be shared with an open license.

\end{document}